\documentclass[5p,twocolumn,times]{elsarticle} 

\usepackage{siunitx}
\usepackage{makecell}
\usepackage{xcolor}
\usepackage{hyperref}
\usepackage{ragged2e, microtype}
\usepackage{subcaption}
\usepackage{pifont}
\usepackage{amssymb}
\usepackage{float}
\usepackage{array}
\usepackage{colortbl}
\usepackage{multirow}
\usepackage{multicol}
\usepackage{amsmath}
\usepackage{algpseudocode}
\usepackage{algorithm}

\newcommand{\cmark}{\ding{51}}%
\newcommand{\xmark}{\ding{55}}%

\usepackage{etoolbox}
\usepackage{soul} 

\newtoggle{finalPaper}

\setstcolor{red}

\toggletrue{finalPaper} 

\iftoggle{finalPaper} {
	\newcommand{\addtxt}[1]{#1}
	\newcommand{\change}[2]{#2}
	\newcommand{\rmvtxt}[1]{}
 	\newcommand{\addtable}[0]{\color{black}}}
{
	\newcommand{\addtxt}[1]{\textcolor{red}{#1}}
	\newcommand{\change}[2]{\st{#1}\textcolor{red}{#2}}
	\newcommand{\rmvtxt}[1]{\st{#1}}
    \newcommand{\addtable}[0]{\color{red}}}

\begin{document}

\title{\LARGE \bf
FederatedTrust: A Solution for Trustworthy Federated Learning}

\author[1]{Pedro Miguel {S\'anchez S\'anchez}}

\author[2]{Alberto {Huertas Celdr\'an}\corref{cor1}}

\author[2]{Ning Xie}

\author[3]{G\'er\^ome Bovet}

\author[1]{Gregorio {Mart\'inez P\'erez}}

\author[2]{Burkhard Stiller}

\address[1]{Department of Information and Communications Engineering, University of Murcia, Murcia 30100, Spain}

\address[2]{Communication Systems Group (CSG), Department of Informatics (IfI), University of Zurich UZH, 8050 Zürich, Switzerland}

\address[3]{Cyber-Defence Campus, armasuisse Science \& Technology, 3602 Thun, Switzerland}

\cortext[cor1]{Corresponding author.
Email address: huertas@ifi.uzh.ch (A. Huertas Celdr\'an)}

\begin{abstract}

\change{With the ever-widening spread of the Internet of Things (IoT) and Edge Computing paradigms, centralized Machine and Deep Learning (ML/DL) have become challenging due to existing distributed data silos containing sensitive information. The rising concern for data privacy is promoting the development of collaborative and privacy-preserving ML/DL techniques such as Federated Learning (FL). FL enables data privacy by design since the local data of participants are not exposed during the creation of the global and collaborative model. However, data privacy and performance are no longer sufficient, and there is a real necessity to trust model predictions. The literature has proposed some works on trustworthy ML/DL (without data privacy), where robustness, fairness, explainability, and accountability are identified as relevant pillars. However, more efforts are needed to identify trustworthiness pillars and evaluation metrics relevant to FL models, as well as to create solutions computing the trustworthiness level of FL models. Thus, this work analyzes the existing requirements for trustworthiness evaluation in FL and proposes a comprehensive taxonomy of six pillars (privacy, robustness, fairness, explainability, accountability, and federation) with notions and more than 30 metrics for computing the trustworthiness of FL models. Then, an algorithm called FederatedTrust has been designed according to the pillars and metrics identified in the previous taxonomy to compute the trustworthiness score of FL models. A prototype of FederatedTrust has been implemented and deployed into the learning process of FederatedScope, a well-known FL framework. Finally, four experiments performed with different configurations of FederatedScope using the FEMNIST dataset under different federation configurations (participant numbers, participants selection rate, training rounds, and differential privacy) demonstrated the usefulness of FederatedTrust when computing the trustworthiness of FL models.}{The rapid expansion of the Internet of Things (IoT) and Edge Computing has presented challenges for centralized Machine and Deep Learning (ML/DL) methods due to the presence of distributed data silos that hold sensitive information. To address concerns regarding data privacy, collaborative and privacy-preserving ML/DL techniques like Federated Learning (FL) have emerged. FL ensures data privacy by design, as the local data of participants remains undisclosed during the creation of a global and collaborative model. However, ensuring data privacy and performance alone is insufficient since there is a growing need to establish trust in model predictions. Existing literature has proposed various approaches on trustworthy ML/DL (excluding data privacy), identifying robustness, fairness, explainability, and accountability as important pillars. Nevertheless, further research is required to identify trustworthiness pillars and evaluation metrics specifically relevant to FL models, as well as to develop solutions that can compute the trustworthiness level of FL models. This work examines the existing requirements for evaluating trustworthiness in FL and introduces a comprehensive taxonomy consisting of six pillars (privacy, robustness, fairness, explainability, accountability, and federation), along with over 30 metrics for computing the trustworthiness of FL models. Subsequently, an algorithm named FederatedTrust is designed based on the pillars and metrics identified in the taxonomy to compute the trustworthiness score of FL models. A prototype of FederatedTrust is implemented and integrated into the learning process of FederatedScope, a well-established FL framework. Finally, five experiments are conducted using different configurations of FederatedScope (with different participants,  selection rates, training rounds, and differential privacy) to demonstrate the utility of FederatedTrust in computing the trustworthiness of FL models. Three experiments employ the FEMNIST dataset, and two utilize the N-BaIoT dataset considering a real-world IoT security use case.}

\end{abstract}

\begin{keyword}
Trustworthy Federated Learning \sep \addtxt{Trust Assessment} \sep \change{Federation Trustworthiness}{AI Governance} \sep Privacy \sep Robustness \sep Fairness \sep Explainability \sep Accountability
\end{keyword}

\maketitle

\section{Introduction}
\label{sec:intro}

The last decade has been a revolutionary time for Artificial Intelligence (AI) \cite{ai_achievements}. IBM Watson, ImageNet, or AlphaGo were some of the first successful AI solutions that defined the path towards the recent ChatGPT, DALL·E 2, or Tesla Autopilot, among many others. This journey has allowed Machine and Deep Learning (ML/DL) models to learn how to play, see, speak, paint, drive, and do many other things, almost like humans. Traditionally, the AI hype has been focused on achieving ever-higher accuracy and performance. However, performance is no longer sufficient. In the last few years, we started to hear more mishaps and situations in which wrong AI-based decisions negatively affect human lives. Some examples are i) ML/DL-based systems supporting judges in pretrial recidivism scoring racially biased~\cite{CynthiaRudin2019}, ii) ML/DL models of autonomous vehicles not prepared nor trained for uncommon fatalities~\cite{MuhammadUzair2021}, or iii) AI-powered chatbots giving wrong answers to straightforward questions and problems~\cite{chatpgt}. These situations erode the trustworthiness of AI and raise concerns about Responsible AI (RAI)~\cite{dignum2019responsible}. 

\change{Trustworthy AI is an emerging concept towards RAI that embraces several existing terms such as explainable AI (XAI)~[6], ethical AI~[7], robust AI~[8], or fair AI~[9], among others}{Trustworthy AI is an emerging concept towards RAI that embraces several existing terms such as explainable AI (XAI), ethical AI, robust AI, or fair AI, among others \cite{trustworthy_ai_from_principles_to_practice}}. In \change{2019, the European Commission published the ethics guidelines for AI}{2021, the European Commission proposed the AI Act \cite{madiega2021artificial}} with high-level foundations, principles, and requirements that AI systems should fulfill to be trustworthy~\cite{eu_guidelines}. According to these guidelines, three main foundations should be met throughout the AI system life cycle. First, AI should be lawful and comply with existing regulations. Second, it should ensure adherence to ethical principles. Last but not least, AI should be robust from technical and social perspectives. Under these three foundations, respect for human autonomy, prevention of harm, fairness, and explainability are four ethical principles that must be respected by trustworthy AI systems. Finally, the European Commission translated these principles into the following seven requirements to achieve trustworthy AI: i) human agency and oversight, ii) technical robustness and safety, iii) privacy and data governance, iv) transparency, v) diversity, non-discrimination, and fairness, vi) societal and environmental wellbeing, and vii) accountability. In parallel to the European Commission, researchers have also developed specific approaches and techniques that AI systems should adopt to be trustworthy. In this context, the systematic reviews on Trustworthy AI conducted in \cite{trustworthy_ai_pillars, trustworthy_ai_from_principles_to_practice} identified robustness, privacy, fairness, explainability, and accountability as the five key pillars of trustworthy AI.

Trustworthiness is a critical aspect influencing AI, but nowadays, it is not the only one, and data privacy and protection are also highly demanded by our society. In this context, new laws and regulations have been drawn in response to this necessity. The General Data Protection Regulation (GDPR) in the European Union and the California Consumer Privacy Act (CCPA) in the state of California (USA) are well-known examples of new data protection regulation~\cite{huertas2020protector}. As expected, these changes affect AI systems since most ML/DL models are trained with data belonging and maintained by different stakeholders in different silos. Therefore, to deal with the challenge of preserving data privacy in AI, Federated Learning (FL) \cite{google_fl} was proposed in 2016 by Google as a decentralized ML paradigm. FL builds collaborative models between the federation members while keeping sensitive data within the premises and control of each participant. In summary, FL is one solution to data silo and fragmentation issues caused by the new legislation that prohibits the free sharing of data and forces data to be maintained by isolated data owners \cite{fl_lecture}. \rmvtxt{FL is highlighted as a key player in the present and future of AI since its global market size is projected to be USD 210 million by 2028 [61].}

\change{The relevance of FL drives academic and industrial actors to consider the trustworthiness of FL model predictions.}{In summary, trustworthiness is critical in FL to address privacy concerns, maintain model integrity, secure the aggregation process, encourage participant cooperation, enable accountability and auditing, and build user trust. By upholding these principles, FL can unlock the potential for collaborative and privacy-preserving ML in various domains while maintaining the highest standards of trust and privacy protection.} \change{To better understand the magnitude of trustworthiness in FL, a comparison with centralized ML/DL has to be made. Like traditional centralized ML/DL, FL is subject to common risks of algorithmic bias, adversarial attacks, data privacy breaches, and reliability issues, among others. However, unlike centralized ML/DL, FL involves different and more stakeholders, actors, information exchanges, communication infrastructures, and attack surfaces that must be considered to assess the trustworthiness of FL. In addition, FL presents new challenges regarding architectural designs, opportunities for privacy-preserving standards, and new perspectives on fairness and explainability beyond the underlying ML/DL models~[13]. In this context, the literature lacks work studying and analyzing trustworthiness pillars and metrics specifically relevant to the context of FL. Furthermore, there is a lack of tools, algorithms, or solutions assessing the trustworthiness level of heterogeneous FL models. Last but not least, there is a need for solutions dealing with trustworthy AI that can be transparently integrated into existing FL frameworks to provide not only traditional performance metrics but also metrics dealing with trustworthy FL.}{Trustworthiness in federated learning (FL) must be examined in comparison to centralized ML/DL. Both approaches share risks such as algorithmic bias, adversarial attacks, privacy breaches, and reliability issues with centralized ML/DL. However, FL introduces additional complexities due to its diverse stakeholders, actors, information exchanges, communication infrastructures, and attack surfaces, necessitating an assessment of trustworthiness. FL also presents unique challenges related to architectural designs, privacy-preserving standards, fairness, and explainability beyond the ML/DL models~\cite{fl_advances_and_opp}. Currently, there is a scarcity of research on trustworthiness pillars, metrics specific to FL, and tools to assess the trustworthiness of FL models. Moreover, there is a need for solutions that seamlessly integrate into existing FL frameworks to compute the trustworthiness of FL models.}

To cover the previous literature gaps, the work at hand presents the following contributions:

\begin{itemize}
    \item The creation of a \addtxt{novel} taxonomy with the most relevant pillars, notions, and metrics to compute the trustworthiness of FL models. To create such a taxonomy, \change{pillars, notions, and metrics}{crucial aspects} used to evaluate the trustworthiness of classical and federated ML/DL models were studied, analyzed, and compared. More in detail, the following six pillars \addtxt{and more than 30 metrics} were identified as the main building blocks of the taxonomy: privacy, robustness, fairness, explainability, accountability, and federation. \change{Each pillar represents a dimension of trust broken down into different notions per pillar, which are then quantified by more than 30 metrics}{All pillars present novel metrics compared to the literature and the one called Federated is novel}.

    \item The design and implementation of FederatedTrust, an algorithm quantifying the trustworthiness of FL models based on the pillars, notions, and metrics presented in the proposed taxonomy. \addtxt{FederatedTrust computes global and partial trustworthiness scores by aggregating metrics and pillars in a dynamic and flexible manner depending on the validation scenario.}  \change{The algorithm}{A prototype of the algorithm} has been implemented \change{as a prototype}{in Python (available in \cite{federatedTrust})} and deployed in a well-known FL framework called FederatedScope. \change{Considering the inner workings of the FederatedScope framework, the prototype algorithm is a Python library (publicly available in [14]) that can be imported into FederatedScope (and other FL frameworks) to generate the inputs needed for the trust evaluation process.}{Then, three experiments classifying hand-written digits in a cross-device FL context using the FEMNIST dataset were performed to assess the trustworthiness of FL models. Experiments introduced differences in the number of participants in the federation, training rounds, sample rates, and countermeasures against attacks. Finally, two experiments leveraged the N-BaIoT dataset to show how different design choices can impact the trustworthiness score in a cybersecurity use case.}

    \item \rmvtxt{The validation of the FederatedTrust prototype in FederatedScope through four experiments classifying hand-written digits in a cross-device FL context using the FEMNIST dataset. Experiments introduced differences in the number of participants in the federation, training rounds, sample rates, and countermeasures against attacks. For all experiments, the trustworthiness values of the FL models were computed (together with the other performance evaluation results) and compared to identify the factors impacting the trustworthiness level of FL.}

\end{itemize}

The remainder of this paper is structured as follows. Section~\ref{sec:related} contains findings from the literature review on trustworthy FL. Section~\ref{sec:pillars} identifies and presents a detailed analysis of the following six trustworthy FL pillars and their metrics: robustness, privacy, fairness, explainability, accountability, and federation. Section~\ref{sec:framework} presents the design detail of the proposed algorithm, while Section~\ref{sec:implementation} focuses on its implementation and deployment on a real FL framework. Section~\ref{sec:validation} validates the algorithm in a use case and presents results from the performed experiments. Finally, Section~\ref{sec:conclusions} provides conclusions and future work.

\section{Related Work}
\label{sec:related}

This section reviews existing solutions focused on trustworthy FL and well-defined pillars relevant to trustworthy AI, such as robustness, privacy, fairness, explainability, and accountability \cite{trustworthy_ai_pillars}. It is important to mention that a large body of literature on trustworthy centralized ML/DL has emerged in recent years. However, trustworthy FL is a nascent research field\rmvtxt{ that requires more effort}. 

FedEval \cite{chai2020fedeval} is the closest solution to the one proposed in this paper because it combines several aspects relevant to trustworthy AI. More in detail, FedEval is an open-source framework for FL systems that evaluates the accuracy, communication, time efficiency, privacy, and robustness of FL models to compute their trustworthiness level. \rmvtxt{FedEval was designed as a benchmark system with a built-in evaluation model.} Regarding accuracy, it compares the performance of FL and the centralized training\rmvtxt{ to determine whether the FL model has achieved better or worse accuracy}. The communication metric relies on the number of communication rounds and the total amount of data transmission during training. The time efficiency metric measures the overall time needed for getting a converged model\rmvtxt{ and the time needed for sub-modules in the model}. The privacy metric \change{is calculated by implementing}{considers} state-of-the-art inference attacks and their impact\rmvtxt{ in the framework, checking the model accuracy and reviewing attack results}. Finally, robustness metrics compute the performance \addtxt{of different aggregation mechanisms }under Non-IID data. Another solution focused on quantifying the trustworthiness of AL models is presented in \cite{celdran2023framework, huertas:2022:ritual}. The authors propose an extensible, adaptive, and parameterized algorithm to quantify the trustworthiness level of supervised ML/DL models with tabular data according to their robustness, explainability, fairness, and accountability. The main limitation of this work is that it is not suitable for FL models. 

Privacy is the central point of FL since its main objective is to protect data privacy among the federation participants. Therefore, it is crucial to preserve and quantify data privacy effectively to trust FL model predictions. In this context, several works and techniques can be categorized into three main families: i) encryption-based, ii) perturbation-based, and iii) anonymization-based. In the encryption-based category, \cite{dong2020eastfly} designed a privacy-preserving protocol against a semi-honest adversary by combining Ternary Gradients with secret sharing and homomorphic encryption. \cite{secure_agg} designed a secure aggregation by leveraging secure multiparty computation to perform sums of model parameter updates from individual users’ devices\rmvtxt{ in a secure manner}. In the perturbation category,\rmvtxt{ the global differential privacy scheme has been widely used in many FL methods. In this sense,} \cite{dp_health} demonstrated that global differential privacy offered a strong level of privacy when protecting sensitive health data in an FL scenario. In addition, \cite{client_dp} proposed a procedure using differential privacy to ensure that a learned model does not reveal whether a client participated during decentralized training. Unlike perturbation-based techniques, anonymization-based techniques can provide privacy defense without compromising data utility. In this category, to measure privacy in FL, \cite{liu2021quantitative} proposed a novel method to approximate the mutual information between local gradient updates and batched input data during each round of training.

Although FL provides a first level of data protection by not sharing training data, the FL paradigm is still vulnerable to adversarial attacks affecting data privacy (inference attacks) and model performance (poisoning attacks)~\cite{fl_attack_taxonomy}. Therefore, robustness is an important pillar to consider in trustworthy FL. To improve the FL robustness against adversarial attacks affecting data privacy and model performance, the authors of \cite{fl_attack_taxonomy} proposed the usage of differential privacy, robust aggregation, and outlier detection as primary defenses. \cite{dp_robust} showed empirical evidence that differential privacy could defend against backdoor attacks and mitigate white-box membership inference attacks in FL. \cite{munoz2019byzantine} introduced Adaptive Federated Averaging (AFA), a Byzantine-robust FL algorithm that detects and discards bad or malicious client updates at every iteration by comparing the similarity of the individual updates to the one for the aggregated model. From a similar perspective, \cite{outlier_backdoor} proposed Robust Filtering of one-dimensional Outliers (RFOut-1d), a new FL approach resilient to model-poisoning backdoor attacks. Finally, regarding mechanisms and metrics able to evaluate the robustness of FL, the literature has focused on methods to quantify robustness once the model is trained. Some examples of these metrics are loss sensitivity, empirical robustness, or CLEVER score \cite{jankovic2022empirical}.

Fairness is another essential pillar for trustworthy FL, as multiple parties contribute data to the model training and are eventually rewarded with the same aggregated global model. In this sense, \cite{fairness_aware_survey} provided a survey and overview of fairness notions adopted in FL-based solutions. The notions include: i) accuracy parity, which measures the degree of uniformity in performance across FL client devices; ii) selection fairness, which aims to mitigate bias and reduce under-representation and never representation; and iii) contribution fairness, which aims to distribute payoff proportionately to the contributions of clients. Apart from that, the authors of \cite{gifair} proposed GIFAIR-FL. This framework imposed group and individual fairness to FL settings by penalizing the spread in the loss of clients to drive the optimizer to fair solutions. FairFL \cite{fairfl} is another solution that facilitated fairness across all demographic groups by employing a Multi-Agent Reinforcement Learning-based scheme. This approach solved the fair classification problem in FL by enforcing an optimal client selection policy on each client. The authors of \cite{selection_fairness} proposed a long-term fairness constraint that considered an expected guaranteed chosen rate of clients that the selection scheme must fulfill. Finally, \cite{fairness_data_valuation} proposed the Completed Federated Shapley Value (ComFedSV) to evaluate data owners' contributions in FL based on solving a low-rank matrix completion problem.

Even with active research on XAI, there are still challenges specifically for FL models, as most client data are private and cannot be read or analyzed. In this context, some explainability methods, such as feature importance, reveals underlying feature information from other parties. For horizontal FL models, since clients share the same feature space, the authors of \cite{shap_fl} suggested that predictions could be explained by calculating the Shapley value of each feature using the definition provided by \cite{molnar2020shapley}. For vertical FL models, their work proposed a variant version of SHAP \cite{shap} by combining the participant features into individual united feature space. Therefore, participants do not get information about the features of other participants. Another solution called EVFL was proposed in \cite{chen2022evfl}, where authors presented a credible federated counterfactual explanation method to evaluate feature importance for vertical FL models and minimize the distribution of the counterfactual and query instances in the client party.

Even though FL models are promising regarding data privacy, they require transparency and accountability, as in the case of classical centralized ML/DL models. In this sense, IBM introduced the Accountable FL FactSheet framework \cite{ibm_fl_factsheets} that instruments accountability in FL models by fusing verifiable claims with tamper-evident facts. The framework requires different actors, like the project owner, data owner, or aggregator, to log claims about the various processes occurring during the FL training lifecycle. They also expanded the IBM AI FactSheet 360 \cite{ibm_factsheet_360} project to account for the complex model compositions of FL. Finally, \cite{desai2021blockfla}, \cite{mugunthan2020blockflow} and \cite{awan2019poster} incorporated blockchain and smart contracts to add different auditing and accountability mechanisms to FL models by leveraging the immutability and decentralized trust properties of blockchain.

\begin{table*}[htb]
\centering
\scriptsize
\caption{Comparison of Related Work}
{\addtable \begin{tabular}{ccccccc}
\hline
\textbf{Solution (year)}     & \textbf{FL} & \textbf{Privacy} & \textbf{Robustness} & \textbf{Fairness} &\textbf{Explainability} &\textbf{Accountability} \\ 
\hline
\hline

\cite{dong2020eastfly} (2020), \cite{secure_agg} (2017)  & \cmark  & Encryption & \xmark & \xmark & \xmark & \xmark \\
\cite{dp_health} (2019), \cite{client_dp} (2017)  & \cmark  & Perturbation & \xmark & \xmark & \xmark & \xmark \\
\cite{liu2021quantitative} (2021)  & \cmark  & Anonymization & \xmark & \xmark & \xmark & \xmark \\

\cite{fl_attack_taxonomy} (2020)  & \xmark  & \xmark & Poisoning and Inference & \xmark & \xmark & \xmark \\
\cite{munoz2019byzantine} (2019), \cite{outlier_backdoor} (2021)  & \xmark  & \xmark & Poisoning  & \xmark & \xmark & \xmark \\
\cite{jankovic2022empirical} (2022) & \xmark  & \xmark  & Assesment & \xmark & \xmark & \xmark\\

\cite{fairness_aware_survey} (2021), \cite{fairness_data_valuation} (2022) & \cmark  & \xmark  & \xmark & Clients Contributions & \xmark & \xmark \\
\cite{gifair} (2021), \cite{fairfl} (2020), \cite{selection_fairness}  (2020) & \cmark  & \xmark  & \xmark & Client Selection & \xmark & \xmark \\

\cite{shap_fl} (2019), \cite{shap} (2017), \cite{chen2022evfl} (2022) & \cmark  & \xmark  & \xmark & \xmark & Feature Importance & \xmark\\

\cite{ibm_fl_factsheets} (2022) & \cmark  & \xmark  & \xmark & \xmark & \xmark & FactSheet\\
\cite{desai2021blockfla} (2021), \cite{mugunthan2020blockflow} (2020), \cite{awan2019poster} (2019) & \cmark  & \xmark  & \xmark & \xmark & \xmark & Blockchain\\

\cite{chai2020fedeval} (2020) & \cmark  & Inference Evaluation & Aggregation & \xmark & \xmark & \xmark \\
\cite{huertas:2022:ritual} (2022), \cite{celdran2023framework} (2023) & \xmark  & \xmark & Poisoning & Data Distribution & Features and Algorithms & Factsheet\\
FederatedTrust (this work) & \cmark  & Perturbation & Poisoning and Inference & Federation & Algorithms and Statistics & Factsheet \\

\hline
\end{tabular}}
\label{table:rel}
\end{table*}

\addtxt{\tablename~\ref{table:rel} compares the solutions covering each one of the FL trustworthiness pillars.} In conclusion, this section has reviewed the work done in each dimension or pillar relevant to trustworthy AI. As can be seen, there is a lack of solutions quantifying the trustworthiness level of FL models by combining the different pillars identified by related work. Most solutions focus on isolated pillars and improving the pillar aspect instead of assessing or quantifying its status, which is needed before deploying countermeasures. In addition, there is no solution dealing with aspects related to the architectural design of the federation.

\section{Trustworthy FL: Pillars, Notions, and Metrics}
\label{sec:pillars}

\rmvtxt{Despite the research done in the field of trustworthy AI, there is a lack of work focused on assessing the trustworthiness of FL models.} This work identifies, introduces, and explains for the first time (to the best of our knowledge) the most relevant pillars, notions, and metrics for trustworthy FL. Additionally, it proposes a novel pillar called Federation, which \change{is not considered in the current literature}{has not been considered in the literature}. This pillar captures complex compositions and designs of FL architectures to compute their trustworthiness. 

\begin{figure*}[ht!] 
\includegraphics[width=\textwidth]{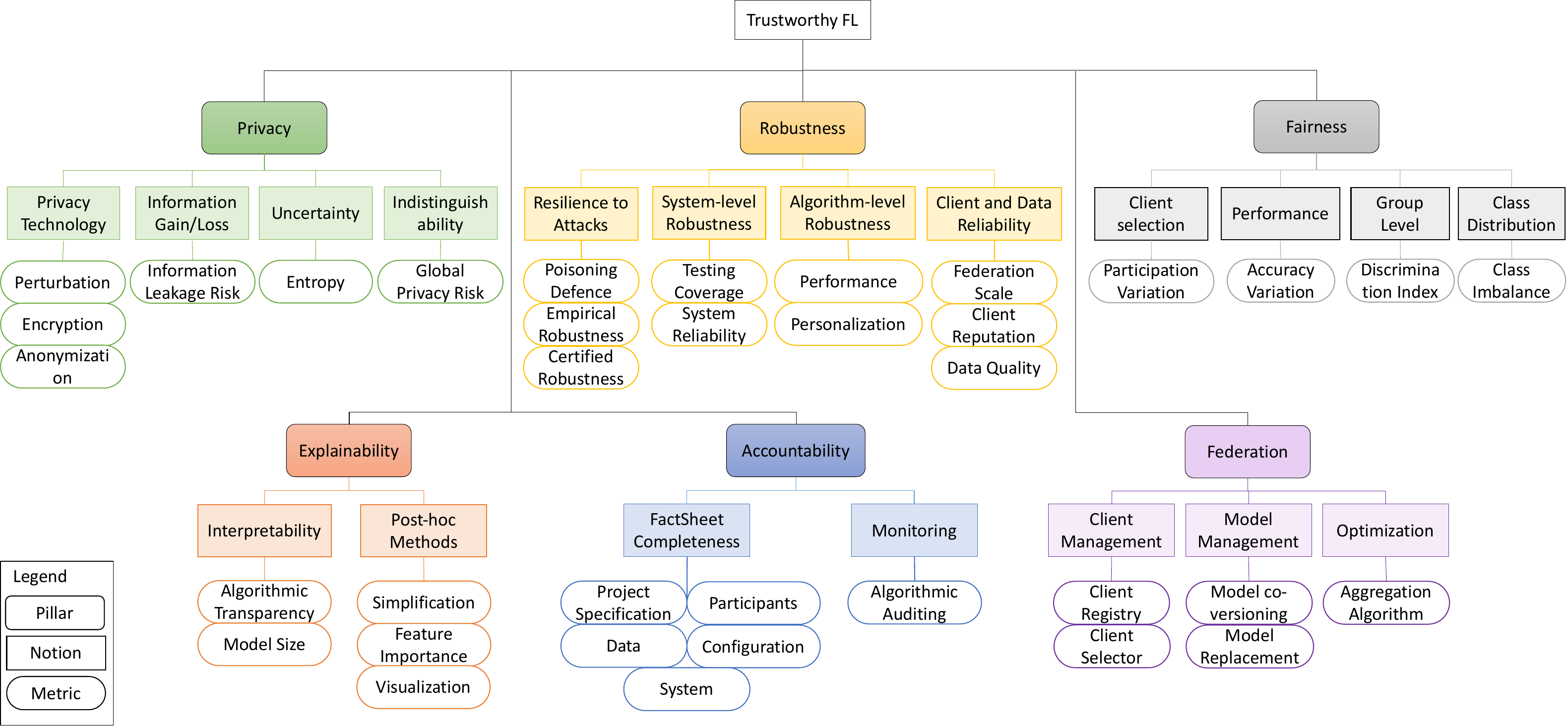}
\caption{Trustworthy FL Taxonomy}
\label{fig:taxonomy}
\end{figure*}

\figurename~\ref{fig:taxonomy} presents a taxonomy describing the pillars, notions, and metrics relevant for trustworthy FL. Under each pillar, the major aspects defining it are grouped into notions. Under each notion, the specific metrics that can be calculated to quantify the level of utility towards trustworthiness are defined. This taxonomy serves as the baseline for evaluating and assessing the trustworthiness level of FL models.

\subsection{Privacy} \label{pillar:privacy}

Data privacy is the most prominent driving force for the development of FL. Therefore, FL models must preserve data privacy within their lifecycle to gain participants' trust. Even though FL already elicits a degree of data privacy by definition, assumptions have to be made about the integrity of the multiple actors and parties making up the federation. If participants are honest, but the aggregating server is honest-but-curious, prevention of information leakage from model parameter exchanges needs to be in place. If all federation members are honest-but-curious, then the prevention of information leakage should focus on secure communication. Moreover, information can still be leaked by malicious attacks from outside. 

To cover these aspects, the first notion of this pillar is the usage of privacy-preserving approaches to add resilience to privacy attacks. The second focuses on metrics measuring information gain/loss based on the information leakage risk derived from the FL process. Finally, two additional notions arise from the probability of knowledge inference from the client updates.

\textbf{Privacy-preserving approaches}. This notion considers the following main approaches to protect data privacy in FL. 

\begin{itemize}
    \item \textit{Perturbation}. It adds noise to raw data, so the perturbed data are statically indistinguishable from the raw data. The most widely adopted schemes are local and global Differential Privacy \cite{wei2020federated}. The global approach adds noise to the parameters shared after the local model is trained, while the local one adds noise to each client data sample used for training. 
    
    \item \textit{Encryption}. It encrypts the model parameters of each participant before sharing it. The most widely adopted scheme is Homomorphic Encryption, where the aggregation server does not decrypt the parameters to aggregate them into the global model \cite{fang2021privacy}. Another popular scheme is Secure Multiparty Computation (SMC) \cite{li2020privacy}, which allows participants to collaboratively calculate an objective function without revealing their data . 
    
    \item \textit{Anonymization}. The most widely adopted schemes in this approach are k-anonymity and l-diversity. K-anonymity is satisfied if each sample in the dataset cannot be re-identified from the revealed data of at least \(k - 1\) clients \cite{choudhury2020syntactic}. L-diversity extends on k-anonymity so that the sensitive attributes of the samples are protected. 
\end{itemize}

\textbf{Information Gain/Loss}. This notion focuses on measuring the amount of privacy lost by participants or the amount of information gained by adversaries due to leakage or disclosure of information \cite{privacy_metrics}. 

\begin{itemize}
    \item \textit{Information Leakage Risk}. In FL the gradients can carry enough information for adversaries to reconstruct the model or infer original data. H-MINE \cite{liu2021quantitative} is a hierarchical mutual information estimation metric to measure the mutual information between the high-dimensional gradients and batched input data. The amount of leaked information (counting the information items disclosed by a system), relative entropy (measuring the distance between two probability distributions), or mutual information (quantifying the shared information between two random variables) are other methods to compute this metric.    
\end{itemize}

\textbf{Uncertainty}. The uncertainty of data estimation by adversaries makes a difference in the level and effectiveness of a data privacy breach. Under normal circumstances, high uncertainty estimation correlates with high privacy. This notion considers metrics to measure uncertainty, most of which are based on entropy.

\begin{itemize}
    \item \textit{Entropy}. In general, entropy measures the uncertainty in predicting the value of a random variable. In FL, an adversary may be interested in identifying which data samples belong to a particular client or organization participating in the training \addtxt{\cite{zheng2023adaptive}}. Equation~\ref{eq:4} calculates the entropy of \(X\), where \(X\) is a participating client and \(p(x_{i})\) is the estimated probability of this client being the target.

    \begin{equation} \label{eq:4}
    priv_{ENT} \equiv H(X) =  -\sum_{x\in{X}} p(x)\log_{2}p(x)
    \end{equation}

\end{itemize}

\textbf{Indistinguishability}. Some adversaries are interested in distinguishing between two data samples of interest. In general, privacy is high if the adversary cannot distinguish between two outcomes of the model.

\begin{itemize}
    \item \textit{Global Privacy Risk}. It enables indistinguishability in the training data by adding random noise. It is a formal statistical guarantee that any disclosure is equally likely whether a sample is in the dataset or not \cite{liu2016preserving}. Equation \ref{eq:dp} describes the formal differential privacy proof for privacy mechanism. A randomized function \(K\) checks if the output of random variables for two datasets \(D_1\), \(D_2\) that differ at most to some extent (e.g., one row of data), differ by at most \(exp(\epsilon)\):
    
    \begin{equation} \label{eq:dp}
    \begin{split}
    priv_{DP} \equiv \forall{S} \subseteq Range(\textit{K}) : p(K(D_1) \in{S}) \leq \exp{(\epsilon)} \\ \cdot{p(K(D_2 \in{S})}
    \end{split}
    \end{equation}
    
\end{itemize}

\subsection{Robustness} 
\label{pillar:robustness}

Robustness is one of the three foundations of trustworthy AI, together with lawfulness and ethics, as defined by the European Commission AI ethics guidelines \cite{eu_guidelines}. AI systems must be technically robust to ensure that they are not vulnerable to malicious use or bring harm to humans. In this sense, the literature has considered three different notions, and this work proposes a new one to assess the robustness of FL models. According to the literature~\cite{trustworthy_ai_from_principles_to_practice}, FL models must be resilient to adversarial attacks adding perturbations or erroneous inputs. Secondly, the hardware and software of participants training and deploying FL models must be robust to avoid cyberattacks \cite{lo2021systematic}. Thirdly, FL algorithms performance and customization must be reliable and robust \cite{kang2020reliable}. Last but not least, this work proposes adding client and data reliability as a novel notion since reliable clients and data increase the probability of robust and reliable FL models. More details of each notion and its metrics are provided below.

\textbf{Resilience to Attacks}. FL models are susceptible to poisoning attacks affecting the model training and its robustness. Poisoning attacks can be categorized into data poisoning and model poisoning. In data poisoning, the integrity of the training data is compromised. Common methods are flipping or permuting the labels and inserting backdoor patterns or perturbations into the training data. Model poisoning attacks have a broader range, and the goal is to manipulate the training procedure. In FL, this could be gradient manipulation or model update poisoning attack, which is performed by corrupting the updates of a participant directly or during model exchanges. Therefore, evaluation of this notion is usually performed by first checking whether the FL model is equipped with any defense mechanism and, if so, verifying the model defense capabilities against representative attacks empirically.

\begin{itemize}
    \item \textit{Poisoning Defense}. It focuses on providing defense mechanisms against poisoning attacks. On the one hand, \textit{Byzantine-resilient Defense} is a popular defense mechanism where various robust aggregation methods have demonstrated their effectiveness in detecting malicious client updates and reducing their impact \cite{fl_advances_and_opp}. On the other hand, \textit{Outlier Detection} explicitly identifies and denies negative influence as a more proactive defense against poisoning attacks. Existing approaches include rejecting updates with too large error rates, measuring the distribution of parameter updates, or looking for dormant neurons that are not frequently activated \cite{wu2020mitigating}.
    
    \item \textit{Empirical Robustness}. If small changes in the input data cause significant output deviations, adversarial perturbations can be used to generate undesired outcomes. It can be measured by implementing a model poisoning attack, a typical poisoning attack altering some local data (data poisoning), or the gradients (model poisoning) \cite{lyu2020privacy}. A mathematical explanation of how a model replacement attack works is provided by \cite{wu2020mitigating} and illustrated as follows. It is assumed that at least one compromised client could apply the backdoor patterns to perform a model replacement attack. Equation \ref{eq:1} describes how a poisoned model update is generated. Where \(w_{t}\) is the current model and \(N\) the number of all clients. The global model \(w\) at time \(t + 1\) is an averaged mean of model updates from \(N\) clients at time \(t + 1\), and the goal is to replace the global model \(w\) at \(t + 1\) with the attacker's model \(x_{atk}\):

    \begin{equation} \label{eq:1}
    x_{atk} = w_{t+1} = w_t + \frac{1}{N} \sum_{i=1}^{N} (x_{t+1}^{i} - w_t)
    \end{equation}

    Now, \(x_{t+1}^{m}\) is the update from the malicious client \(m\) at time \(t+1\), then rearranging Equation \ref{eq:1}, we have:

    \begin{equation} \label{eq:2}
    x_{t+1}^{m} = N \cdot x_{atk} - N \cdot w_t - \sum_{i=1}^{N-1} (x_{t+1}^{i} - w_t) + w_t
    \end{equation}

    Assuming \(\sum_{i=1}^{N} (x_{t+1}^{i} - w_t) \approx{0}\) as explained in \cite{bagdasaryan2020backdoor}, we have the attacker's update be simplified as the following:

    \begin{equation} \label{eq:3}
    x_{t+1}^{m} = N \cdot (x_{atk} - w_t) + w_t
    \end{equation}
    
    \item \textit{Certified Robustness}. It defines the least amount of perturbation required for the attacker to succeed (change the model prediction). In other words, a model is certifiably robust for an upper-bounded amount of perturbations. The CLEVER (Cross Lipschitz Extreme Value for nEtwork Robustness) metric \cite{clever} using the local Lipschitz constant for neural networks is one of the most well-known metrics. CLEVER is an attack-agnostic derivation of the universal lower bound on the minimal distortion required for a successful attack.
\end{itemize}

\textbf{System-level Robustness}. This notion deals with hardware and software robustness and must be considered in production environments with FL models where proper software development and deployment standards are needed. Testing coverage and system reliability are the main metrics for this notion.


\begin{itemize}
    \item \textit{Testing Coverage}. It guarantees that clients adhere to the federation requirements such as broadcasting messages to distributed clients, client selection, and model aggregation. This metric can be implemented using different methodologies, such as robust system delivery, ranging from code review, unit testing, integration testing, system testing, and acceptance testing \cite{gitiaux2021aura}. 

    \item \textit{System Reliability}. It deals with the probability and the duration of time of failure-free operation \cite{reliability_metrics}. It is measured based on error, timeout, and dropout rates. The error rate is normally calculated as the number of failures over a given amount of time. The maximum timeout measures the time the server should wait to receive client model updates. Finally, the dropout rate is the number of clients leaving the federation (to speed up convergence, optimize resources, or due to errors) divided by the total amount of clients. A high dropout rate can also indicate a less reliable FL system. 
\end{itemize}

\textbf{Algorithm-level Robustness}. This notion deals with the performance and generalization of FL algorithms. Performance is widely used to showcase how good an ML/DL model is. However, good performance does not necessarily imply generalization. Generalization is a major challenge in FL because each client has different local data heterogeneity, and the aggregated global model might not be able to capture the data pattern of each of them. Non-IID data \cite{fl_non_iid} may cause severe learning divergence to parametric models. Therefore, it is desirable to have metrics to measure the performance, and generalization of FL \cite{lo2022architectural}. 

\begin{itemize}
    \item \textit{Performance}. In FL, there are two approaches to measuring performance. One is reserving a set of test or validation data on the server side to be used for global model evaluation. The other is by evaluating test accuracy at each client's device and aggregating the test accuracy. This metric considers these two aspects. 

    \item \textit{Personalization}. Several methods have been proposed for this category of metrics. Regularization is one of the personalized FL approaches aiming to minimize the disparity between global and local models. Multi-tasking Learning is a method where multiple learning tasks are solved simultaneously. This benefits FL models since multiple organizations participating in the FL models can train their personalized models to achieve better performance. For example, MOCHA \cite{multi_task} generates separated but related models on local client devices using data of related tasks. Clustering is another personalization approach where clients are allocated into clusters based on their similarity \cite{sattler2020clustered}. Personalized Layer is a metric for neural network models that introduces custom layers for each client in the model. FedPer \cite{fedper} is an example that uses the base layers as the shallow layers and personalized layers as the deep layers while keeping everything else the same as the baseline FedAvg algorithm.
\end{itemize}

\textbf{Client and Data Reliability}. This novel notion of robustness deals with client and data reliability. In this sense, metrics like federation scale, clients' reputation, and data quality are essential for robust and trustworthy FL models.

\begin{itemize}
    \item \textit{Federation Scale}. The number of clients impacts the reliability of the system. In FL, the number of clients determines the number of devices, network connections, and model parameters that must be considered to train FL models. The higher the number of clients, the network stability, computation power, and availability of clients the higher and reliability of the FL system \cite{li2021survey}.

    \item \textit{Client Reputation}. In \cite{kang2020reliable}, a novel reputation metric was proposed, and a subjective logic model was used to calculate the reputation score for each client interaction. After each training iteration, the server uses a poisoning attack detection scheme and the elapsed time to determine if the local update from the client is reliable. Reliable updates are treated as positive interactions and improve the reputation value, and vice versa. The client is treated as malicious and unreliable when the reputation value is below a threshold.

    \item \textit{Data Quality}. It compares, in each training round, the local updates with the current global model to see if the new local updates are better or worse than the global model \cite{pejo2020quality}. If a round of training at one client improves the performance of the global model, then the data from that client has a high-quality score.

\end{itemize}
\subsection{Fairness} 
\label{pillar:fairness}

One primary source of unfairness in AI is coming from the data. In FL, different clients might contribute with heterogeneous amounts and quality of data. When data are not representative of the wider federation, participant selection bias is propagated into the model. The selection bias can be manifested by feature distribution skew or label distribution skew \cite{fl_advances_and_opp}, both of which are major challenges in FL. Therefore, Client Selection Fairness is the first notion of this pillar. Apart from that, in fair AI \cite{fair_ai}, fairness is broken down into group-level and individual-level fairness. Group-level fairness means that members of a particular group should not be subject to discrimination. Individual-level fairness means that similar individuals should receive similar treatment regardless of their membership group. These requirements do not change when moving to FL. Therefore, the Group-level Fairness notion deals with the group-level aspects, and the Performance Fairness and Class Distribution notion covers individual-level ones. More in detail, Performance Fairness ensures that each client's reward should be proportional to their data contribution. Finally, Class Distribution looks at the label imbalance in the dataset of each participant.

\textbf{Client Selection}. Usually, in FL, only a fraction of clients are selected to participate in the training process of each round. In practice, there are several criteria for selection, such as availability, network speed, computation power, or battery level, among others. In the worst case, clients located in regions with low network speed or models with weaker computation power could never get selected and represented in the training data. Therefore, under-representation is a source of selection bias that could lead to unfair model outcomes. 

\begin{itemize}
    \item \textit{Participation Variation}. In statistics, the Coefficient of Variation (CV) measures how far the data values from a set are dispersed from the mean. This metric analyzes the distribution of participation rates among all clients. With similar clients, the more dispersed the distribution of participation rate, the less fair the client selection mechanism is, and vice versa \cite{alvi2022federated}. Equation \ref{eq:cv} calculates the CV in the participation rate, where $\sigma$ represents the standard deviation, and $\mu$ represents the average of the participation rate.

    \begin{equation}
    \label{eq:cv}
        CV = \frac{\sigma}{\mu}
    \end{equation}
\end{itemize}

\textbf{Performance}. Even though performance fairness already exists in AI, in FL, another grouping needs to be considered, which is client-level fairness. \cite{fair_resource} suggests a definition that a model provides a more fair solution to the FL learning objective on the clients if the performance is more uniform than that of another model.

\begin{itemize}
    \item \textit{Accuracy Variation}. This metric considers the test accuracy as a representation of performance. More in detail, the aggregated global model and test data from each client are used to measure the test accuracy. The more uniform the test accuracy among the clients, the more fair the model performance is.
\end{itemize}

\textbf{Group-level}. Evaluating and mitigating demographic bias in FL is more difficult than in centralized learning. First, raw training data, labels, and sensitive demographic information of each participant cannot be revealed. Second, in centralized learning, all the training data can be analyzed and pre-processed to balance the class distribution before training. In contrast, different clients pre-process their data locally in FL, so additional mechanisms are needed to adjust the global data distribution in a secure and protected manner.

\begin{itemize}
    \item \textit{Discrimination Index}. The discrimination index metric \cite{fairfl} measures the difference in the F1 score between a particular demographic group ($\sigma$) and the rest of the population. The metric value falls between \([-1, 1]\), where the ideal discrimination index should be as close to 0 as possible. Calculating this index globally would reveal sensitive attributes and statistics of the demographic group in the client data. Equation \ref{eq:disc} calculates the discrimination index, where $F1(X^+_\nu)$ represents the F1 score of all the samples from the protected group and $F1(X^-_\nu)$ represents the F1 score of the rest of the samples. 

    \begin{equation}
        \label{eq:disc}
        \Phi_\sigma = F1(w(X^+_\sigma))-F1(w(X^-_\sigma))
    \end{equation}
    
\end{itemize}

\textbf{Class Distribution}. Analyzing the class distribution of the training data used in an ML/DL model provides insight into whether the data samples are selected properly to reflect a fair representation of the wider group. In theory, this should also apply to FL models, except that in practice, this often requires access and analysis of the raw training data, which goes against data privacy. However, in \cite{class_imbalance_reduction}, an estimation scheme was proposed to reveal the class distribution without accessing raw data. Furthermore, secure aggregation can also be considered for aggregating class distribution information among clients.

\begin{itemize}
    \item \textit{Class Imbalance}. Two approaches can be used to evaluate the class imbalance. One is the estimation scheme using a well-balanced auxiliary dataset and the gradients of a neural network model \cite{class_imbalance_reduction}. Another general way to get the class imbalance information in FL is to ask every client to submit their class distribution to the server. The secure aggregation method combines all the class distributions into one unified distribution. Then the coefficient of variation of the class distribution can be used to calculate the level of variation of the class sample sizes to determine the class imbalance.
\end{itemize}

\subsection{Explainability} 
\label{pillar:explainability}

AI guidelines demand that AI processes should be transparent, with the capabilities and the purpose of AI systems openly communicated and decisions explainable to those directly and indirectly impacted. Transparency is often expressed as interpretability which is often wrongly mistaken as interchangeable with explainability. Interpretability is the first notion of this pillar and can be described as a passive characteristic of a model referring to the level of understandability for humans. In contrast, explainability is the ability to provide a description of the technical process of an AI system. Interpretable models can be explained by analyzing the model itself, but for non-interpretable models, Post-hoc Methods can enhance their interpretability, being the second notion of this pillar. For FL, since ML/DL models are also used in the training process, the requirement of explainability for the algorithmic model also applies. However, the privacy constraint makes it difficult to access and analyze raw data or protected model features directly.

\textbf{Interpretability}. This notion combines the algorithm transparency and the model size to evaluate the FL model interpretability. As already identified in the literature, some ML/DL models are interpretable by design, and some are not. In addition, model size is also widely used as a metric of interpretability. More details about these two metrics are provided below.

\begin{itemize}
    \item \textit{Algorithmic Transparency}. A model is considered transparent if it is understandable by itself. This definition could be very subjective to different levels of intellectual grasp. However, algorithmically transparent models must first be fully explorable through mathematical analysis and methods. Then the assessment considers model complexity (in terms of the number of variables and interactions) and decomposability (in terms of the interpretability of each component of the model) \cite{XAI}. \addtxt{In summary, the recognized interpretable models are linear regression, logistic regression, decision trees, decision rules, k-nearest neighbors (KNN), and Bayesian models. Even within the interpretable models, the level of interpretability slightly varies. The non-interpretable models include tree ensembles, support vector machines (SVM), multi-layer neural networks (MNN), convolutional neural networks (CNN), and recurrent neural networks (RNN).}
    
    \item \textit{Model Size}. Different algorithms have diverse definitions of model size. For instance, it could be the number of decision rules, the depth of a decision tree, the number of features in a linear/logistic regression model or the number of trainable parameters in a neural network \cite{quantification_xai}. The larger the model size, the harder it is to understand and explain the causal relationship between input and output. 
\end{itemize}

\textbf{Post-hoc Methods}. The three most common Post-hoc Methods are simplification, feature importance, and visualization \cite{XAI}. \change{Simplification means reducing the complexity of a non-interpretable model into a simpler surrogate model that is more easily interpreted and understood. The feature relevance explanation method measures each feature influence, relevance, or importance in the prediction output, hoping to explain what aspect of the training population causes certain model behavior. Finally, visualization methods provide visual representations of AI models to facilitate better understanding by humans. Below more details about the two last metrics are provided.}{This notion is complementary to the previous and contributes to assessing the explainability of interpretable and non-interpretable FL models.}

\begin{itemize}
    \item \textit{Simplification}. The idea of simplification lies in reducing the number of architectural elements or parameters in a model. One of the main techniques applied for model simplification is knowledge distillation \cite{gou2021knowledge}. 

    \item \textit{Feature Importance}. Most model explanation methods can be directly used for horizontal FL because all participants share the full feature space in their local data \cite{shap_fl}. However, exposing feature information to the server for calculating feature importance score is not ideal. For vertical FL, methods like SHAP cannot be directly used because parties do not share the full feature space. The variant version of SHAP for Vertical FL combines one party's features into a federated feature space when referenced by another party for the feature importance calculation \cite{shap_fl}.

    \item \textit{Visualization}. A lifecycle dashboard that visualizes information from the server, starting from client registration to training, validation, and deployment, was proposed in \cite{fl_xai_dashboard}. The dashboard shows which clients participated in which round of training and the current status of the model.
    
\end{itemize}

\subsection{Accountability} 
\label{pillar:accountability}

Accountability is another of the seven critical requirements of Trustworthy AI defined by the EU guidelines \cite{eu_guidelines}. The first main notion about accountability is FactSheet Completeness. IBM Research was the first to propose a document called FactSheet in charge of recording facts about the overall ML/DL pipeline \cite{ibm_factsheets}. Another important notion of accountability is Monitoring. Even with complete and detailed documentation, every participant has to make an effort to ensure that FL models are built strictly following the intended architecture, development, and deployment processes.


\textbf{FactSheet Completeness}. IBM extended the FactSheet approach to enable accountability in FL \cite{ibm_fl_factsheets}. The accountable FL FactSheet template is a comprehensive document that contains meta-information about the project, participants, data, model configurations, and performance. Since FL is more complicated in architecture and more privacy-preserving, the FactSheet should contain information about the additional layer of configurations and avoid sensitive information about participating clients. Below more details about the aspects considered in FactSheets are provided. 

\begin{itemize}
    \item \textit{Project Specification}. This section of the FactSheet documents the project overview, purpose, and background. The overview explains what the project is about. The purpose details the goals, and the background elaborates on the relevant information and knowledge motivating the project.
    
    \item \textit{Participants}. It contains the participants of the FL process. The template considers the participants' names and their organization unit names for identity verification. 
    
    \item \textit{Data}. It documents the information regarding the data used in the FL process. Two aspects are included: data provenance and pre-processing procedures. Data provenance helps trace the data origin and flow to access validity and even reputation. Pre-processing steps can tell how the raw data have been handled before training the model. 

    \item \textit{Configuration}. It deals with the information about the FL model configuration. First, it contains the type of optimization algorithm and the ML/DL model. Then, it indicates the global hyper-parameters the aggregator uses, for instance, the number of rounds, the maximum timeout, and the termination accuracy. Lastly, it contains the local hyper-parameters used by the trainer at each client, such as the learning rate and the number of epochs.
    
    
    
    \item \textit{System}. This FactSheet section documents the system information for the learning process. It includes the average time spent on training, the model size, the model upload, and the download speed in bytes. This information indicates the number of resources expected to be utilized.
\end{itemize}

\textbf{Monitoring}. 
For AI systems, algorithmic auditing is a range of approaches to auditing algorithmic processing systems. The evaluation of the metrics under this notion uses a checklist-based approach, verifying if the FL system employs any external or internal algorithmic auditing.

\begin{itemize}
    \item \textit{Algorithmic Auditing}. This metric can be implemented in different ways. For example, it could be functional testing, performance testing, user acceptance testing, etc. It could also be system anomaly or attack monitoring. Some organizations even invite or hire external hackers to find vulnerabilities as a measure of monitoring. The SMACTR framework \cite{auditing} is a good option for a more systematic approach. It is an internal auditing framework with five stages: scoping, mapping, artifact collection, testing, and reflection. Each stage yields a set of documents that form a comprehensive audit report. 
\end{itemize}

\subsection{Federation} 
\label{pillar:design}

The major management challenges of FL deal with communication, efficiency, resource limitation, and security. In this context, it is very challenging to coordinate the learning process of thousands of clients while ensuring model integrity and security. Global models might converge slowly due to heterogeneous client data. Inconsistent clients, networks, and limited resources might cause clients to drop out, and training failures could impact model quality. In conclusion, although there is active research in FL algorithms, there is still a lack of research and guideline on the architectural design of FL systems. In this sense, the main notions of this pillar are Client and Model Management, which considers how client and model information is administrated in the system, and Optimization Algorithm, which may impact the model performance and robustness.


\textbf{Client Management}. This notion proposed a Client Registry, where participants can register themselves for training, and a Client Selector to filter eligible clients for training. More details are provided below.

\begin{itemize}
    \item \textit{Client Registry}. It enables the system to manage client connections and track the status of all client devices. The proposed design pattern maintains the client registry in the central server for the client-server architecture. The server sends a request for information along with the initial local model to the clients when they first connect to the system. The information requested includes device ID, connection up and down time, or device computation power storage. 
    
    \item \textit{Client Selector}. It optimizes resource usage and reduces the risk of client dropout and communication latency. The proposed design pattern also maintains the client selector in the central server where the selection occurs. Before each round of training, the client selector actively selects a certain number of clients for the training according to predefined criteria to reduce convergence time and optimize the model performance.
\end{itemize}

\textbf{Model Management}. In a distributed learning process like FL, multiple rounds of training and aggregation of models generate numerous local model updates and aggregated global models during the process. Without recording the local and global intermediary models, there is neither traceability nor fallback when something goes wrong in the training process. A model co-versioning registry and replacement can help trace the model quality and improve system accountability. \addtxt{Blockchain has been proposed by recent works of the literature to populate individuals models in an immutable and transparent manner, mitigating some attacks affecting FL \cite{baniata2022machine}.}

\begin{itemize}
    \item \textit{Model Co-versioning}. It aligns the local model versions with their corresponding aggregated global models. It can be a registry where local model versions are stored and mapped to the associated global models. With this registry, model updates and aggregations do not always have to be synchronous because the server can refer to the mapping to perform asynchronous aggregations. Another advantage is that it allows early stopping if a model converges before the specified number of rounds.
    
    \item \textit{Model Replacement}. It detects the global model performance dropping below a certain threshold level. For that, it compares the global modal performance in all clients to see if the performance degradation is a global event. The new global model training task is triggered if the degradation is global and persistent.
\end{itemize}

\textbf{Optimization}. According to the goal and the context of FL models, the choice of an optimization algorithm can impact the model performance. Therefore, various studies have conducted performance benchmarking of FL optimization algorithms \cite{nilsson2018performance}. This benchmarking comparison serves as a reference for this metric.

\begin{itemize}
    \item \textit{Aggregation Algorithm}. FedAvg is considered the baseline aggregation algorithm, but some other optimization algorithms have been proposed as an extension for various purposes \cite{nilsson2018performance}. \addtxt{In addition, the aggregation can be done both in a centralized or decentralized manner. Decentralization reduces network bottleneck, single point of failure, and trust dependencies, while increasing network overhead in some cases \cite{beltran2022decentralized}. This metric considers these two aspects to evaluate the trustworthiness level of the aggregation task.}
    
\end{itemize}

\section{FederatedTrust Algorithm Design}
\label{sec:framework}

This section details the design of \textit{FederatedTrust}, the algorithm proposed to quantify the trustworthiness level of FL models according to the pillars, notions, and metrics presented in Section~\ref{sec:pillars}. \addtxt{To the best of our knowledge, it is the first attempt to evaluate the trustworthiness of FL models.}

\change{Before presenting the algorithm details, it is important to mention that this work assumes that the central server, aggregating the clients' models, is honest and maintained by a trusted system administrator}{Prior to delving into the specifics of the algorithm, it is crucial to mention that this work operates under the assumption that the central server, which aggregates the models of the clients, is honest, maintains data integrity, and is overseen by a dependable system administrator}. Therefore, the server does not maliciously interfere with the trust calculation process. In addition, the following functional requirements (FR), non-functional requirements (NF), and privacy constraints (PC) have been considered before designing and implementing the proposed algorithm.

\begin{enumerate}
    \item[FR-1:] Each of the six trustworthy FL pillars must be represented in the algorithm, meaning that at least one metric from each pillar must be considered in the final score.
    \item[FR-2:] The final trustworthiness score must be a combination of the trustworthiness scores from all notions and pillars.
    \item[NF-1:] The algorithm should add minimal computation overhead and complexity to the server, participants, and FL model.
    \item[NF-2:] The algorithm should be modular and configurable. 
    \item[PC-1:] The algorithm must not store any sensitive data from the FL model.
    \item[PC-2:] The algorithm must not leak or share any sensitive data from clients, the server, and the FL model with third parties.
    \item[PC-3:] The metrics calculations can occur at the client's local devices, the central server, or collaboratively between both. 
    \item[PC-4:] When metrics are calculated collaboratively between clients and the server, the computation should be performed securely and privately if the individual client metrics contain sensitive information.
\end{enumerate}

\begin{figure*}[htpb!] 
\centering
\includegraphics[width=0.8\textwidth]{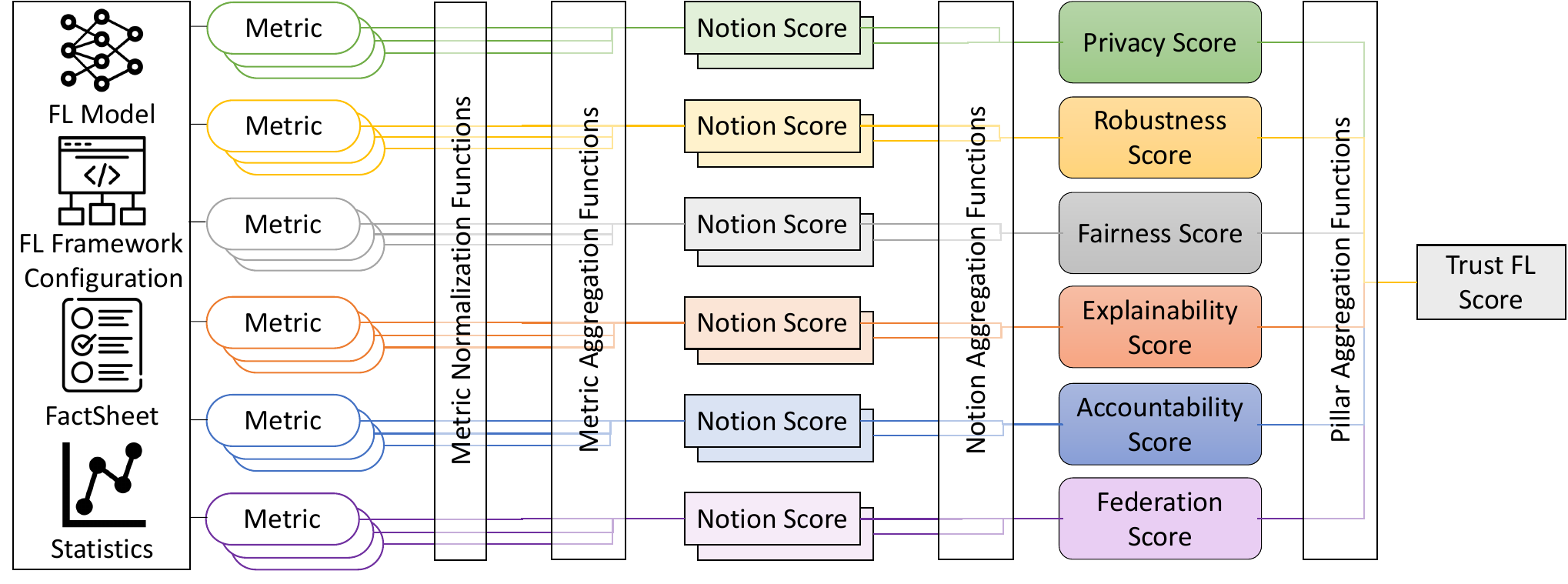}
\caption{Design of the FederatedTrust Algorithm}
\label{fig:algo_design}
\end{figure*}

Once assumptions and requirements are defined, \figurename~\ref{fig:algo_design} shows the overview of the \text{FederatedTrust} algorithm design. First, the proposed algorithm considers the following input sources to compute the trustworthiness of FL models.

\begin{itemize}
    \item \textbf{FL Model}. The FL model trained in a collaborative and privacy-preserving way between the federation participants. This input contains information about the model configuration and model personalization.
    
    \item \textbf{FL Framework Configuration}. The configuration parameters of the tool implementing the protocol needed to train and evaluate the FL model. This input contains information about the number of clients, the client selection mechanisms, the aggregation algorithm, and the model hyperparameters. 
    
    \item \textbf{FactSheet}. \change{A document containing relevant information about the training process, the federation, and its participants. More in detail, this input contains information about the FL project overview, its purpose and background, the data provenance, pre-processing techniques, and differential privacy mechanisms.}{As mentioned in Section~\ref{sec:pillars}, It provides essential details for the accountability of the training process, federation, and the individuals involved. Specifically, it encompasses information about the overview of the task to solve, data origin, techniques used for pre-processing, and the incorporation of differential privacy mechanisms.}
    
    \item \textbf{Statistics}. The statistical information extracted from the training dataset of each client and its model performance. It does not contain sensitive information. This input contains information about the client class balance, client test performance loss, client test accuracy, client clever score, coefficient of variation for client feature importance, coefficient of variation for client test accuracy, coefficient of variation for clients participation rate, coefficient of variation for client class imbalance, client average training time, average model size, average upload bytes, and average download bytes.
\end{itemize}

These input sources are used to compute the metrics indicated in Section~\ref{sec:pillars}, which output values are then normalized to have a common range of values. It is important to mention that each metric can consider different input sources and can be calculated in a different phase of the FL model creation process (pre-training, during-training, or post-training) and by a different actor of the federation (client or server). These details regarding when and who computes each metric are provided later. Once the normalized outputs of metrics are calculated, they are weighted and aggregated to compute one score per notion (see Section~\ref{sec:pillars} for more information about notions). Each pillar has one or more notions calculated according to predefined but configurable weights per metric. Therefore, the same process is repeated to obtain the pillar scores from weighting and aggregating notion scores. Finally, the final trust score of the FL model is a configurable combination of the pillar scores. The implementation details of the previous steps are provided in Section~\ref{sec:implementation}.

\begin{figure}[t!] 
\centering
\includegraphics[width=\columnwidth]{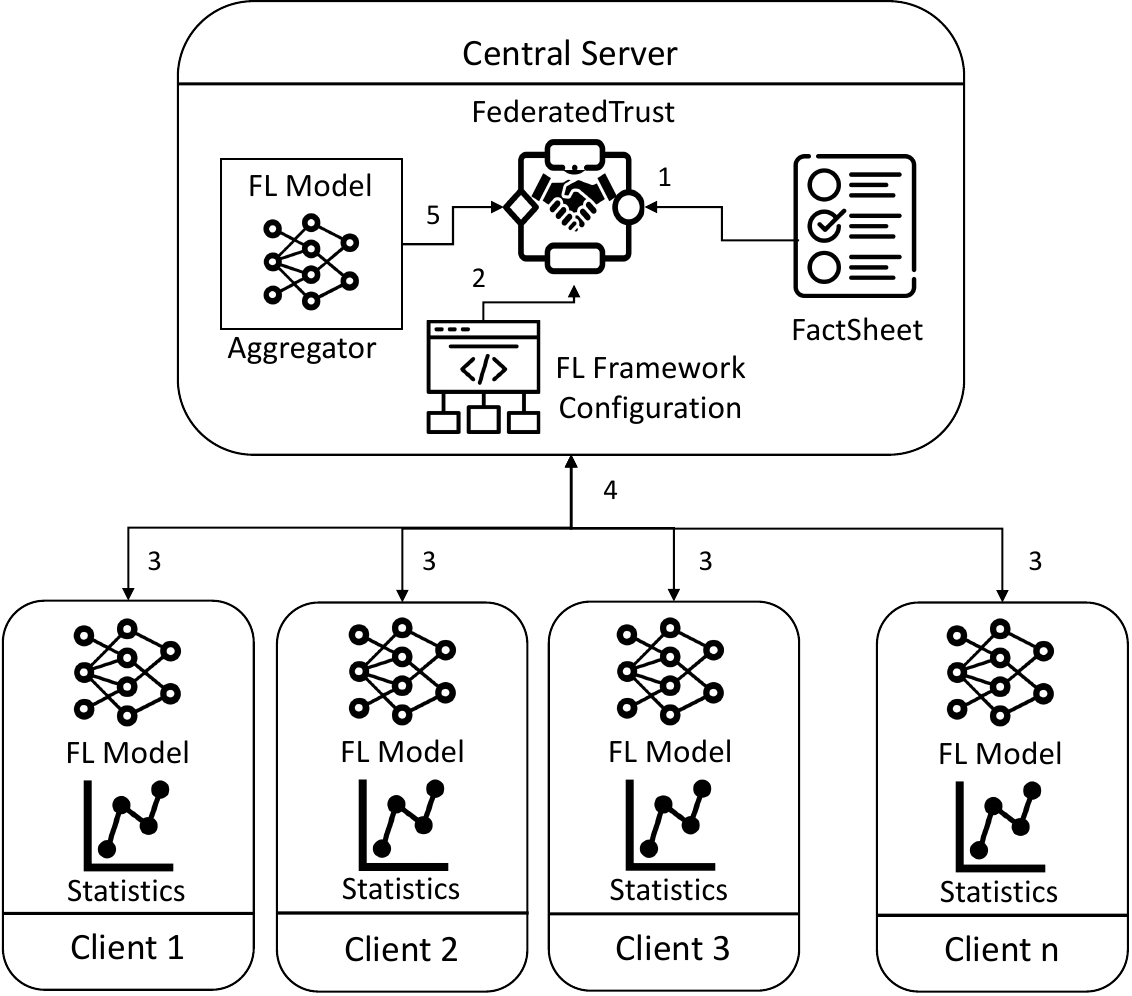}
\caption{Overview of the Integration of FederatedTrust into an FL Framework}
\label{fig:scenario}
\end{figure}

As indicated, some metrics of the algorithm are calculated during the training phase of the FL model. Therefore, it is necessary to integrate the proposed algorithm into an FL framework in charge of creating FL models. In this context, \figurename~\ref{fig:scenario} shows the interactions between the main actors involved in the computation of the trustworthy level of FL models. As it can be seen, the central server of the federation hosts i) the \textit{Aggregator}, in charge of combining the parameters of the clients' models to create the FL model, ii) the \textit{FactSheet}, accounting the most important aspects of the FL project and participants federation, iii) the \textit{FL Framework Configuration}, detailing aspects of the framework, and iv) the \textit{FederatedTrust} algorithm, computing the FL model trust score. To compute the trust score, during the pre-training phase, the content of the FactSheet and the FL framework configuration (detailed in Section~\ref{sec:implementation}) is sent to the FederatedTrust algorithm (steps 1 and 2 in \figurename~\ref{fig:scenario}) and metrics depending on them are calculated as previously explained. Then, the model training process starts with the server sharing the type of model and its characteristics with the clients of the federation (step 3). Once clients locally train their models with their private datasets, they send the models parameters and statistics of their data to the server for aggregation (step 4). At that point, the parameters are aggregated, and the FL Model and statistics (about training datasets and model performance) are sent to the FederatedTrust algorithm for computing metrics during training (step 5). Steps 3, 4, and 5 are repeated until the FL model converges or the training rounds are over. At this point, the training process concludes, and FederatedTrust computes the post-training metrics. \addtxt{Finally, the FederatedTrust algorithm provides a score per pillar and a global one, together with a trustworthiness report}. The detailed version of the previous steps is provided by Algorithm \ref{alg:algorithm}.


\begin{algorithm}[t!]
\scriptsize
\caption{Training a FL model equipped with FederatedTrust}\label{alg:algorithm}
\begin{algorithmic}[1]
\State \textbf{Input:} clients \(N\), sampling size \(m\), a central server \(S\), number of iterations \(T\), initial model \(\bar{w}^{(0)}\), FL Framework configuration \(C\), FederatedTrust \(ft\), FactSheet \(fs\)
\State \textbf{Output:} global score, pillars scores, trustworthiness report
\State \(S\) sends the hashed IDs of all clients \(i \in [N]\), \(C\), and \(fs\) to \(ft\) 
\State \(ft\) creates a map of hashed client IDs to values of 0, representing the initial selection rate
\State \(S\) sends the model metadata to \(ft\)
\State \(S\) request class distribution information from all clients \(i \in [N]\)
\For{clients \(i \in [N]\)}
\State Client \(i\) uses \(ft\) function to calculate the sample size per class of local data
\State \(ft\) creates or updates the class distribution map of hashed labels to sample size
\EndFor
\For{\(t\) = 0 to \(T\)}
\State \(S\) randomly samples \(D^{(t)} \subset{[N]}\) clients with size of \(m\)
\State \(S\) sends the hashed IDs of the selected clients to \(ft\)
\State \(ft\) updates the client selection rate map
\State \(S\) broadcasts the current model \(\bar{w}^{(t)}\) to all clients \(i \in D^{(t)}\)
\For{clients \(i \in D^{(t)}\)}
\State Client \(i\) performs local training with \(\bar{w}^{(t)}\)
\State Client \(i\) sends new model updates \(w_i^{(t+1)}\) back to \(S\)
\State Client \(i\) computes evaluation metrics with local test data and local model \(\bar{w_i}^{\;'}\)
\State Client \(i\) sends the evaluation results back to \(S\)
\EndFor
\State \(S\) performs secure aggregation of all updates into a new global model \(\bar{w}^{(t+1)}\)
\EndFor
\State \(S\) aggregates the evaluation results and sends them to \(ft\)
\State \(ft\) receives the evaluation results and populates them
\State \(S\) asks \(ft\) to evaluate the trustworthiness of the model
\State \(ft\) computes the trustworthiness score and generates a report JSON and print message
\end{algorithmic}
\end{algorithm}

\section{FederatedTrust Algorithm Prototype}
\label{sec:implementation}

This section contains the implementation details of the FederatedTrust algorithm when dealing with a given problem presented as a use case. \rmvtxt{In more detail, first, a well-known and powerful FL framework able to train heterogeneous FL models has been selected. After that, a set of metrics, from the list provided in Section~3, common for different FL use cases, has been implemented in the FederatedTrust algorithm to compute the trustworthiness level of different FL models. Then, a use case where ML models must be trained in a collaborative and privacy-preserving way has been selected. Finally, FederatedTrust has been deployed and integrated into the FL framework to solve the use case.} The implemented prototype is available in~\cite{federatedTrust}.

\subsection{FL Framework Selection and Use Case}

Regarding the selection of the FL framework, after careful comparison and analysis of the most relevant and used frameworks in charge of training diverse FL models (TensorFlow Federated, PySyft, Flower, FLUTE, LEAF, FederatedScope, FedEval, and FedML), FederatedScope~\cite{federatedscope} was chosen as a reference tool. \rmvtxt{FederatedScope is an open-source FL framework developed in 2022 by the Data Analytics and Intelligence Lab (DAIL) of Alibaba DAMO Academy~[73]. FederatedScope employs an event-driven architecture to provide users with great flexibility. It is a comprehensive framework that builds on existing research work~[74] and offers a suite of functionalities for privacy and adversarial attacks.} More in detail, the following functionality lead to the selection of FederatedScope in this work.

\begin{enumerate}
    \item \change{Provides s}{S}tandalone and distributed modes to set up clients experimentally or realistically.

    \item \change{Provides d}{D}ifferential privacy and inference attacks.
    
    \item \change{Provides w}{W}ell-documented evaluation metrics.
    
    \item \addtxt{Data zoo, a suite of well-known federated datasets such as FEMNIST~\cite{leaf}, and Algo zoo, a list of optimization algorithms such as FedAvg, or FedProx.}
    
    \item \change{Includes M}{M}odel zoo, a set of computer vision and language models.
\end{enumerate}

Then, a basic use case was chosen for implementing and testing the deployment of the FederatedTrust algorithm as a proof of concept. In this context, the use case focuses on classifying handwritten digits and numbers in a federated and privacy-preserving way. For that, FederatedScope runs an FL training process in standalone mode with a variable number of clients and one central server over a defined number of iterations. The function to aggregate individual models and create the global FL Model is FedAvg, and the model is a convolutional neural network (CovNet2). The federated datasets that the clients hold locally are from the FEMNIST image dataset of handwritten digits and letters. The dataset has 62 classes (10 digits, 26 lowercase, 26 uppercase), and the images are 28 by 28 pixels. The client selector strategy is random sampling.

\rmvtxt{To compute the trustworthiness level of FL models classifying handwritten numbers in a privacy-preserving and federated fashion, the following metrics were implemented in the FederatedTrust prototype.} 


\subsection{Metrics Selection and FederatedTrust Implementation}

As seen in Section~\ref{sec:pillars}, not all metrics have a standardized way of being computed, available equations, or feasible calculation methods. Therefore, \change{the objective of this section is to implement a}{the objective is to implement a} lightweight prototype with the basic pillars, notions, and metrics that can be calculated in any FL project created with the FederatedScope framework. The list of omitted notions and metrics and the reasons for not including them in the prototype implementation are the following.

\begin{itemize}
    
    \item \textit{Encryption and anonymization (Privacy)}. The usage of these two techniques is not documented in the configuration file of the FederatedScope framework. Furthermore, the implemented prototype considers differential privacy as a perturbation technique to protect data privacy. 

    \item \textit{Information Leakage Risk (Privacy)}. This metric needs to be calculated during every round of the training process by running extra neural networks, which would incur high computation overhead.

    \item \textit{Poisoning Defense (Robustness)}. \change{It is often integrated into the FL training protocol, for example, with robust aggregation. Therefore, it is complicated}{It is challenging} to quantify the usage of the poisoning defense mechanism unless the information is documented in the FL framework, which is not the case in FederatedScope. \addtxt{In addition, another way to verify this metric is to check the certified robustness against poisoning attacks, which is considered in the implementation.}

    \item \textit{Empirical Robustness (Robustness)}. This metric depends on each attack type, and the number of possibilities and complexity is huge. Furthermore, certified robustness could help to cover this metric. 
    
    \item \textit{Testing Coverage and System Reliability (Robustness)}. Neither testing data nor error rates are available in the FederatedScope framework.
    
    \item \textit{Client Reputation (Robustness)}. There is no clear way to measure client reputation in a simulated environment. It requires more inspections of the client data provenance.
    
    \item \textit{Data Quality (Robustness)}. It needs to be executed during every round of the training process, creating a high computation overhead for the FL model training process.

    \item \textit{Discrimination Index (Fairness)}. It requires knowledge of protected sensitive features and it is unclear how this calculation can be done without leaking sensitive information\rmvtxt{between clients and the server}.

    \item \textit{Visualization (Explainability)}. Graphical capabilities to show the explainability of FL models are not included in the FederatedScope framework.

    \item \textit{Algorithmic Auditing (Accountability)}. It is not implemented \change{as the validation scenario uses simulation data,}{because} there are no real end users, and \change{it is assumed no attackers}{no attackers are present} in the environment.

    \item \textit{Client Registry (Federation)}. The FederatedScope framework simulates the clients participating in the federation. Therefore, no client registration is required\rmvtxt{, as in real scenarios}.
    
    \item \textit{Model co-versioning and replacement (Federation)}. Versioning and model replacement trigger functionality is not implemented as the validation scenario is simulated.
\end{itemize}

Once those metrics excluded for the prototype implementation are indicated, \tablename~\ref{table:arch} describes the list of implemented metrics, with descriptions, inputs, output values, calculation moment, and federation actor in charge of computing them.

\begin{table*}[htb]
\centering
\scriptsize
\caption{Metrics Implemented by the FederatedTrust Algorithm Prototype}
\begin{tabular}{p{0.14\textwidth}|p{0.42\textwidth}|p{0.12\textwidth}|p{0.04\textwidth}|p{0.09\textwidth}|p{0.04\textwidth}} 
\hline
\textbf{Metric}      & \textbf{Description} & \textbf{Input} & \textbf{Output} & \textbf{When} &\textbf{Who}\\ 
\hline
\hline
\multicolumn{6}{c}{\textbf{Privacy}} \\ 
\hline
Differential Privacy &  Use of global or local differential privacy as a privacy defense & FactSheet & 0/1 & Pre-training& Server\\
\hline
Entropy &  Uncertainty in predicting the value of a random variable & FL Framework Conf& [0, 1] & Pre-training & Server \\
\hline
Global Privacy Risk &  Maximum privacy risk with differential privacy based on \(\epsilon\) & Client Statistics & \%  & Pre-training & Server\\
\hline
\multicolumn{6}{c}
{\textbf{Robustness}} \\ 
\hline
Certified Robustness & Minimum perturbation required to change the neural network prediction & FL Model & Real & Post-training & Server \\
\hline
Performance &  Test accuracy of the global model & Statistics, FL Model & \% & During-training & Clients\\
\hline
Personalization &  Use of personalized FL techniques &  FactSheet &  0/1 & Pre-training & Server \\
\hline
Federation Scale &  Number of clients representing the scale of the federation & FactSheet &  Integer & Pre-training & Server  \\
\hline
\multicolumn{6}{c}{\textbf{Fairness}} \\ 
\hline
Participation Variation & Uniformity of distribution of participation rate among clients & FL Framework & [0, 1] & Post-training & Server\\ 
\hline
Accuracy Variation & Uniformity of distribution of performance among clients & Client Statistics, FL Model & [0, 1]  & During-training & Clients \\
\hline
Class Imbalance & Average class imbalance estimation among clients & Client Statistics  & [0, 1]   & Pre-training & Clients \\
\hline
\multicolumn{6}{c}{\textbf{Explainability}} \\ 
\hline
Algorithmic Transparency & Interpretability of the model by design & FL Model & [1, 5] & Pre-training & Server \\ 
\hline
Model Size & Model Features dimensionality, depth of decision tree, or number of parameters in neural networks & FL Model & Integer & Post-training & Server \\
\hline
Feature Importance & Average variance of feature importance scores & FL Model & [0, 1]   & Post-training & Server \\
\hline
\multicolumn{6}{c}{\textbf{Accountability}} \\ 
\hline
Project Specification & Project details and purpose  & FactSheet & 0/1 & Pre-training & Server  \\ 
\hline
Participants & Participants number, identifiers, and their organizations  & FL Framework Conf & 0/1  & Pre-training & Server  \\ 
\hline
Data & Contains Data origin and data-preprocesing steps & FactSheet & 0/1  & Pre-training & Server\\ 
\hline
Configuration & Information about the FL model & FL Framework Conf, FactSheet & 0/1  & Pre-training & Server \\ 
\hline
System & Contains training time, FL model size, and network performance & FL Framework Conf, Statistics & 0/1  & Post-training & Server    \\ 
\hline
\multicolumn{6}{c}{\textbf{Federation}} \\ 
\hline
Client Selector & Use of a client selector scheme rather than random selection & FactSheet & 0/1  & Pre-training & Server   \\ 
\hline
Aggregation Algorithm & Selected aggregation function & FL Framework Conf & \% & Pre-training & Server  \\ 
\hline
\end{tabular}
\label{table:arch}
\end{table*}

In order to understand how the trust FL score is computed, it is important to mention that the output value of each metric can have different ranges, as indicated in the metric definitions (\tablename~\ref{table:arch}). Therefore, to combine these values into an understandable trustworthiness score, different functions are used to translate the output values into a normalized score between [0, 1]. The logic of each normalization function is explained in \tablename~\ref{table:metric_scores}. It shows for each metric the type of values it gets, the range or set of possible values, and how these values are mapped into the final output between 0 and 1. Once all metrics outputs are normalized, they are grouped (first by notions and then by pillars) to compute the Trust FL score. The algorithm prototype uses the mean average of all the metrics under a notion to calculate its final score. In the same way, the mean average of the notions of one pillar is employed as the pillar score. Finally, the Trust Score of the entire setup is the mean average of all the pillars. How metrics, notions, and pillars are combined can be changed to different aggregation methods based on the requirements of the environment, giving more importance to some metrics, notions, or pillars. The implementation of the previous life-cycle is done by using the \textit{numpy}, \textit{scipy}, and \textit{sklearn} libraries. More in detail, FederatedTrust is designed as a third-party library.

\begin{table*}[htb]
\centering
\scriptsize
\caption{Normalization of the Metrics Outputs}
\begin{tabular}{p{0.13\textwidth}|p{0.32\textwidth}|p{0.23\textwidth}|p{0.22\textwidth}} 
\hline
\textbf{Metric} &
\textbf{Type} & \textbf{Output} &  \textbf{Normalized Output} \\ 
\hline
\hline

\multicolumn{4}{c}{\textbf{Privacy}} \\ 
\hline

Differential Privacy & True/False & 0/1 &  0/1 \\ 
\hline
Entropy & Uncertainty & [0, 1] &  [0,1] \\ 
\hline
Global Privacy Risk & Percentage & [0, 100] &  [0, 1] \\ 
\hline

\multicolumn{4}{c}{\textbf{Robustness}} \\ 
\hline

Certified Robustness & CLEVER Score & [0, 0.2, 0.4, 0.6, 0.8, 1, 1.2, 1.4, 1.6, 1.8, 2, 2.2, 2.4, 2.6, 2.8, 3, 3.2, 3.4, 3.6, 3.8, 4.0] &  \{0, 0.05, 0.1, ... 1\} \\ 
\hline
Performance & Accuracy & [0, 1] &  [0, 1] \\ 
\hline
Personalization & True/False & 0/1 & 0/1 \\ 
\hline
Federation Scale & Number of clients & [10, $10^2$, $10^3$, $10^4$, $10^5$, $10^6$]&  \{0, 0.2, 0.4, ... 1\} \\ 
\hline

\multicolumn{4}{c}{\textbf{Fairness}} \\ 
\hline

Participation Variation & Coefficient of Variation & [0, 1] &  [0, 1] \\ 
\hline
Accuracy Variation & Coefficient of Variation & [0, 1] & [0, 1] \\ 
\hline
Class Imbalance & Balance Ratio & [0, 1] & [0, 1] \\ 
\hline

\multicolumn{4}{c}{\textbf{Explainability}} \\ 
\hline

Algorithm Transparency & Random Forest, K-Nearest Neighbors, Support Vector Machine, GaussianProcessClassifier, Decision Tree, Multilayer Perceptron, AdaBoost, GaussianNB, Quadratic Discriminant Analysis, Logistic Regression, Linear Regression, Sequential, Convolutional Neural Network  & \{4, 3, 2, 3, 5, 1, 3, 3.5, 3,4, 3.5, 1, 1\} &  \{0, 0.2, 0.4, 0.5, ... 1\} \\ 
\hline
Model Size & Number of Model Parameters & \{1, 10, 50, 100, 500, 1000, 5000, 10000, 50000, 100000, 500000\} &  \{0, 0.1, 0.2, 0.3, 0.4, 0.5, 0.6, 0.7, 0.8, 0.9, 1.0\} \\ 
\hline
Feature Importance & SHAP Importance & [0, 1] &  [0, 1] \\ 
\hline

\multicolumn{4}{c}{\textbf{Accountability}} \\ 
\hline

 Project Specification & True/False & 0/1 & 0/1 \\ 
\hline
Participation & True/False & 0/1 & 0/1 \\ 
\hline
Data & True/False & 0/1 & 0/1 \\ 
\hline
Configuration & True/False & 0/1 & 0/1\\ 
\hline
System & True/False & 0/1 & 0/1 \\ 
\hline

\multicolumn{4}{c}{\textbf{Federation}} \\ 
\hline

Client Selector & True/False & 0/1 &  0/1 \\ 
\hline
Aggregation Algorithm & FedAvg, FedOpt, FedProx, FedBN, pFedMe, Ditto, FedEM & \{0.8493, 0.8492, 0.8477, 0.8548, 0.8765, 0.8661, 0.8479\} &  \{0.8493, 0.8492, 0.8477, 0.8548, 0.8765, 0.8661, 0.8479\} \\ 
\hline

\multicolumn{4}{l}{[] denotes a continuous range and \{\} denotes a list of values}
\end{tabular}
\label{table:metric_scores}
\end{table*}


\section{Experiments}
\label{sec:validation}

\addtxt{This section shows how the FederatedTrust framework can be integrated and employed together with FederatedScope to perform trustworthiness evaluation in FL applications during model generation. The demonstration experiments are organized into two groups. The first set of three experiments using the FEMNIST dataset that show how the number of clients and their configuration impact the trustworthiness. Then, a second set of two experiments leveraging the N-BaIoT dataset about IoT network security, which show how different training configurations impact a real-world use case.}

\subsection{Trustworthiness Scores for Experiments with FEMNIST}

The previously defined setup combining FederatedTrust and FederatedScope was used to perform \addtxt{the following }\change{four}{three} experiments that consist of training FL models that classify hand-written digits using the FEMNIST dataset. \rmvtxt{The details of each experiment are provided below.}

\begin{itemize}
    
    \item \textit{Experiment 1}: The experiment considers a federation of 10 clients with a selection rate of 50\% clients per training round\change{. In other words, 5 randomly selected clients out of the 10 available ones participate in each round}{ (5 out of 10 are randomly selected in each round)}. The training of the FL model runs 5 rounds. \rmvtxt{Finally, the experiment does not use personalization techniques. Therefore, all clients share the same global model.}
    
    \item \textit{Experiment 2}: It considers 50 clients \change{for the federation. In}{and in} every iteration the server randomly selects 60\% clients\rmvtxt{ to participate in the training}. The experiment runs 25 training iterations. \change{As in the previous experiment, no personalization techniques are used. However, t}{T}he main novelty of this experiment lies in the inclusion of differential privacy with \(\epsilon\) of 20.
    
    \item \rmvtxt{Experiment 3: This experiment has a similar configuration to Experiment 2 but with more data privacy due to a lower value of epsilon (6) used in differential privacy. It means that the expected privacy protection would be higher.}
    
    \item \textit{Experiment \change{4}{3}}: This experiment \rmvtxt{is medium-scale among all the previous ones. It}consists of 100 clients with a 40\% client selection rate and 50 rounds of training. The \(\epsilon\) value of differential privacy is set to 6. 
    
\end{itemize}

\change{For all experiments, the same values about the project specifications, data, and participants are extracted. The purpose and the background of the FL project are not specified in the model. Finally, all models used a random sampling client selector and ran the FedAvg aggregation algorithm. Because of that, the federation pillar score does not change between the four experiments.}{In all experimental setups, the same configuration of the project specifications, data, and participants were employed. The specific purpose and background information of the FL project were not explicitly stated within the model. Furthermore, all models implemented a client selection method based on random sampling, and the FedAvg aggregation algorithm was utilized. As a result, the federation pillar score remained constant across all \change{four}{three} experiments.}

\begin{figure*}[htpb!] 
\centering
\includegraphics[width=\textwidth]{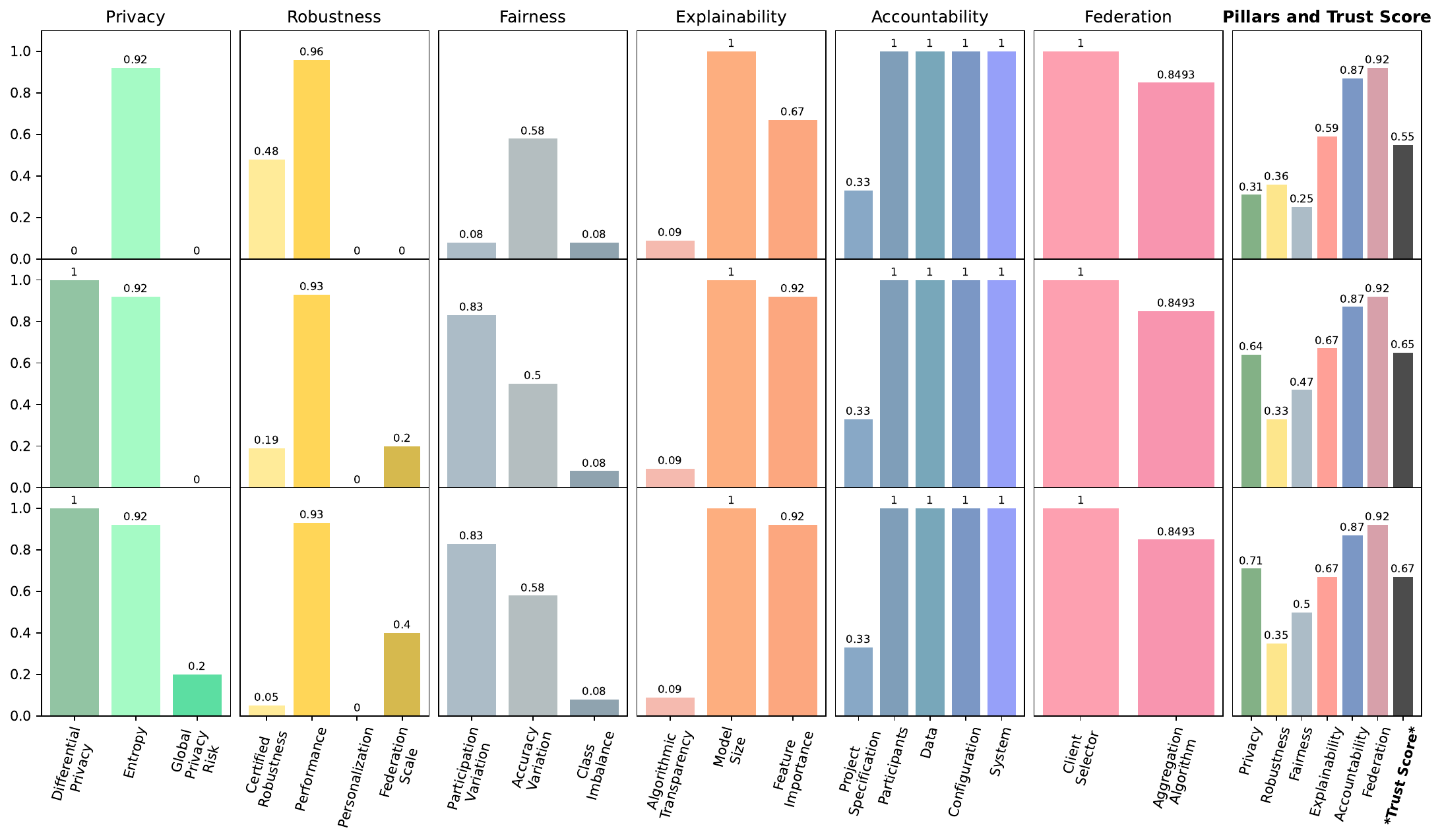}
\caption{\addtxt{Experiment Comparison with FEMNIST Dataset}}
\label{fig:exp}
\end{figure*}

As can be seen in \change{Figure 3 and Figure 4}{\figurename~\ref{fig:exp}}, the trustworthiness score for both \addtxt{first} experiments is 0.56. However, there are differences in terms of pillars and metrics. 

Starting from the privacy pillar, its score increased from 0.31 (Experiment 1) to 0.64 (Experiment 2) because of using differential privacy. However, the effectiveness of the privacy mechanism was not good enough because of the large value of \(\epsilon\) chosen (20). It can be seen in the indistinguishability notion (covered by the global privacy risk metric), with a score of 0 in both experiments. In this context, the larger the \(\epsilon\) value, the less noise was added to the data, and therefore, the higher the probability of being identified by adversaries. Dealing with the fairness pillar, its score from Experiment 1 to 2 was also increased from 0.25 to 0.47. It was mainly due to a significant increment in the selection variation metric (from 0.08 to 0.83), which resulted from the overall increase in the number of clients, the client sampling rate, and the number of rounds. The selection fairness improved with more clients participating and more training rounds. However, the performance variation metric (Fairness pillar) dropped from 0.58 to 0.50 from Experiment 1 to 2. It could also be due to the increased number of clients. More clients with different levels of heterogeneity in their data could influence the generalizability of the global model hence affecting the individual test accuracy at the client level. Furthermore, personalization techniques were not used, so the global model was not adapted to the clients. Regarding the explainability pillar, its global score increased from 0.59 (Experiment 1) to 0.67 (Experiment 2), with an increase in the feature importance metric score from 0.67 to 0.92. It is impacted by the different numbers in terms of clients and the differences between their datasets. Finally, the robustness pillar score dropped from 0.36 (Experiment 1) to 0.33 (Experiment 2). It was mainly due to the decrement in the certified robustness metric (from 0.48 to 0.19) and the performance reduction from 0.96 to 0.93. The drop in the certified robustness metric could be related to the increase in the number of clients and the number of rounds. In theory, more aggregating parties provide more entries and surfaces for adversaries to insert backdoor perturbations for poisoning attacks. There are also higher chances for parties to collude when they are more in number. The higher number of rounds also means that adversaries have more chances to attack. The federation scale metric for both experiments has a similar score since both have less than 50 clients.



\rmvtxt{\subsection{Trustworthiness Scores for Experiments 3 and 4}}

\rmvtxt{Figure 3 and Figure 4 show the trustworthiness score for both experiments with global scores of 0.64 and 0.67, respectively.} Before comparing \change{these two experiments}{Experiments 2 and 3}, it is important to mention that the main difference between \change{Experiments 2 and 3}{them} is the privacy pillar (due to the \(\epsilon\) change). \rmvtxt{The remaining pillar scores are very similar. }Focusing on Experiments 2 and 3, from the robustness pillar perspective, its overall score only increased from \change{0.36 (Experiment 3) to 0.38 (Experiment 4)}{0.33 (Experiment 2) to 0.35 (Experiment 3)}, even though there was a significant increase in the federation score from 0.2 to 0.4. It was mainly due to the less relevance of that metric in the pillar than the rest. Furthermore, the certified robustness score decreased from 0.19 to 0.05. The privacy pillar score increased from \change{0.68 (Experiment 3) to 0.71 (Experiment 4)}{0.64 (Experiment 2) to 0.71 (Experiment 3)} due to the global privacy risk metric. The increase in the number of clients caused an improvement in the indistinguishability notion (covered by the previous metric). Assuming that random guessing was used, in Experiment \change{4}{3} was twice as challenging to guess the correct target among 100 clients compared to 50 clients (used in Experiment \change{3}{2}). In terms of explainability, the score \change{increased from 0.61 (Experiment 1) to 0.92 (Experiment 2) due to the feature importance score, which continued to increase to 0.92 in Experiment 4}{remained constant in Experiments 2 and 3 as no changes were made in the factors impacting these metrics}. \rmvtxt{It might be related to the increase in data samples from more clients and the better model performance. When the model learned the data better, the features had more impact on the prediction outcomes.} Finally, the fairness score was almost identical for both experiments  (0.47 and 0.5). It might be because the ratio between the increase in the number of clients and the increase in the number of rounds was the same, while the clients' sampling rate remained the same as well.


\addtxt{\subsection{Turstworthiness Scores for Experiments with N-BaIoT}}

\addtxt{To perform a more exhaustive comparison, two more experiments were performed. In this case, the N-BaIoT \cite{meidan2018n} dataset was leveraged. It contains benign and attack network traces from nine different IoT devices. In the two experiments done with this dataset, the number of clients is nine, one per IoT device. Besides, the same configuration of project specifications, data, and participants was employed.}

\begin{itemize}
    \item \addtxt{\textit{Experiment 4:} It employs FedAvg as the aggregation algorithm and does not use Differential Privacy. Additionally, the client selection ratio is 70\% and no local data balancing is performed, having a 20/80 ratio between benign and attack labels. Finally, the classification model is a deep neural network with two hidden layers of 50 neurons.}

    \item \addtxt{\textit{Experiment 5:} It leverages Federated Median as aggregation approach and employs local Differential Privacy with \(\epsilon\) equal to 4. The client selection ratio is 90\% and local dataset balancing is applied in each client, leaving a 50/50 balance per class. The neural network, in this case, contains four hidden layers of 100, 80, 70, and 50 neurons.}
    
\end{itemize}

\begin{figure*}[htpb!] 
\centering
\includegraphics[width=\textwidth]{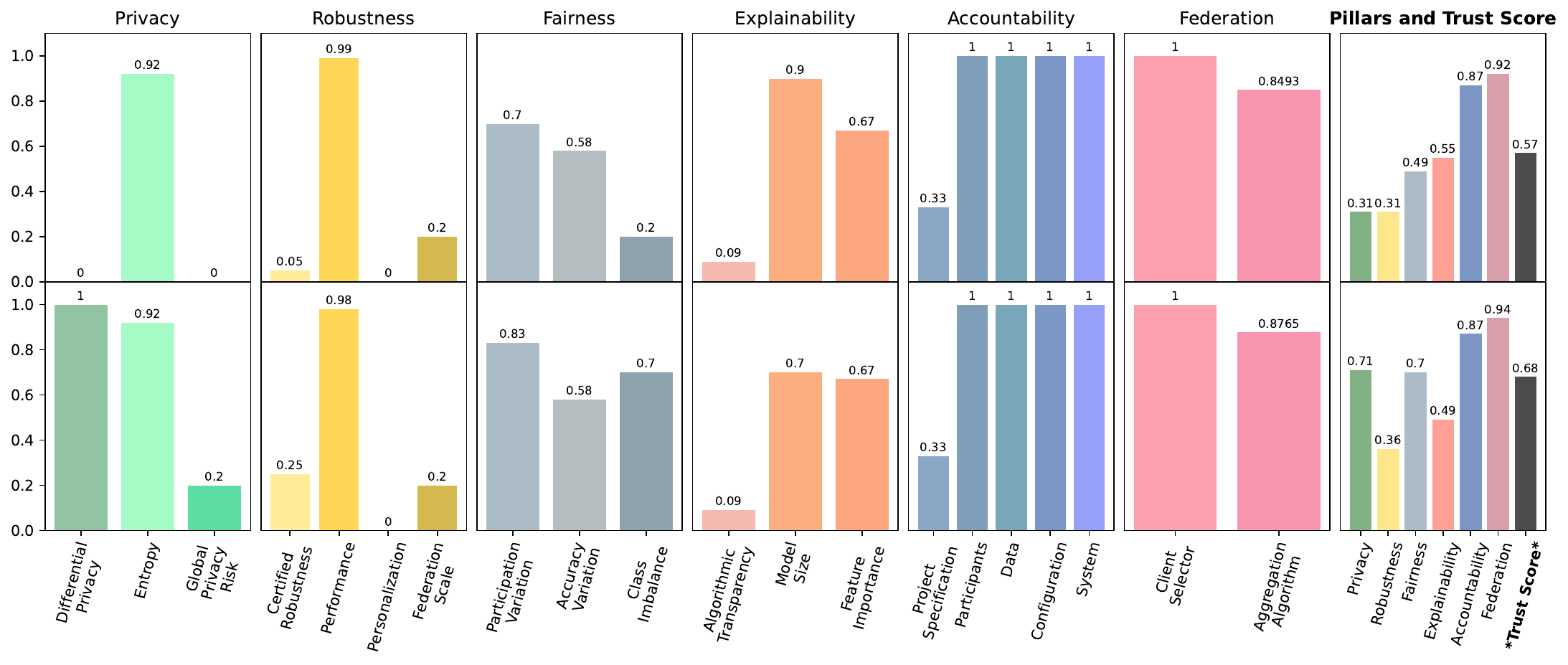}
\caption{\addtxt{Experiment Comparison with N-BaIoT Dataset}}
\label{fig:exp2}
\end{figure*}

\addtxt{\figurename~\ref{fig:exp2} shows the different trustworthiness metrics in Experiment 4 (up) and 5 (bottom). Comparing Experiments 4 and 5, it can be seen that even though the overall performance slightly decreases from 0.99 to 0.98, the final trust score increases from 0.57 to 0.68. The decrease in the performance is caused by the usage of a different aggregation algorithm and the usage of differential privacy, as adding noise to the training data can decrease the final model performance. The increase in the trust score is happening due to the increase in most of the evaluated pillars due to the differences in the experiment configurations. The privacy pillar goes from 0.31 to 0.71 due to the usage of differential privacy and the associated increase (from 0 to 0.2) in the global privacy risk metric. The robustness pillar is also increased even though the overall performance decreases by 0.01. This occurs because the certified robustness increases from 0.05 to 0.25, enhancing the pillar score from 0.31 to 0.36. Similarly, the fairness pillar is increased from 0.49 to 0.7 due to a better client participation ratio and class imbalance management. In contrast, the explainability pillar is decreased from 0.55 to 0.49 because the model size metric goes from 0.9 to 0.7 due to the usage of a larger model in Experiment 5. Finally, the federation pillar is slightly increased from 0.92 to 0.94 thanks to the usage of median aggregation instead of averaging.}

\addtxt{From the comparison of Experiments 4 and 5, it can be seen how a real federated application can leverage configuration optimizations during its design in order to increase its trustworthiness without losing notable performance. Note that more changes could be applied according to the metric to optimize depending on the exact use case.}

\subsection{Limitations}

There are several limitations of the FederatedTrust prototype in terms of quantifying the trustworthiness level of FL models. First, \change{some metrics are easily quantifiable and can represent a component of the trustworthiness level well. For example, certified robustness represents the attacker lower bound, which is directly associated with resilience to attacks. On the other hand, other}{some} metrics, like the federation scale (robustness pillar), often have to be considered with other factors to represent the notion well. For example, the analysis performed in Section~\ref{sec:pillars} shows that, in practice, the client reputation metric is also an essential factor for the client reliability notion. However, it was difficult to quantify the client reputation in the performed experiments. Another example would be the client selection fairness notion. Although the client participation variation metric is easily quantifiable by computing the dispersion of selection rate among the clients, this variation metric alone might not be the best representation of fairness for client selection in FL. \rmvtxt{Increasing the number of clients, the sample rate, and the round of training could easily bring up the participation variation metric, but equality does not necessarily imply equity which is important for true fairness.}

Another limitation pertains to the algorithm scoring \addtxt{and metric aggregation} systems. \rmvtxt{First, to aggregate all the scores into a trustworthiness level between 0 and 1, all the metric values have to go through the metric operations and the scoring functions. }First, the logic of the scoring functions greatly impacts how the trust score of each metric is calculated. Based on the pillar analysis in Section~\ref{sec:pillars}, there were general directions of how every metric should impact the overall trustworthiness level. However, the concrete scoring maps and ranges were created based \change{on subjective understanding and standards used}{knowledge} from other studies\change{. Their generalizability for other}{and their generalizability to other} systems was not fully evaluated. \change{Furthermore, the current metric weighing system does not implement varied weights, meaning that every metric has the same weight under one notion, and every notion has the same weight when aggregated into the final trustworthiness level. The weighing system could be enhanced by further analysis of the trade-offs between pillars and metrics to produce a more balanced trustworthiness level of FL models.}{Second, although FederatedTrust provides a flexible approach to decide the importance of each metric and pillar, selecting optimum values is challenging because it presents a trade-off between pillars. In this context, multi-objective optimization techniques \cite{saini2021multi} could be integrated into FederatedTrust to optimize the weights selection according to the needs of each scenario where the algorithm is deployed.}

\addtxt{Finally, some challenges dealing with resource consumption, data leakage, governance and compliance, and scalability might appear when deploying FederatedTrust in real scenarios. For example, implementing metrics evaluated by FederatedTrust requires careful design and consumes computational resources that might not be available in resource-constrained devices. In addition, it is critical to guarantee data privacy while computing the trustworthiness score of FL models \cite{alzubi2022cloud}. FederatedTrust must also adhere to regulatory and legal requirements, so maintaining transparency, accountability, and auditability is a challenge. Finally, implementing scalable and distributed mechanisms (like Blockchain) to populate metrics outputs while maintaining privacy, integrity, and trust poses a significant challenge~\cite{zarour:2020:blockchain}.
}

\section{Conclusions and Future Work}
\label{sec:conclusions}

\change{This work has studied and analyzed the pillars, notions, and metrics relevant to trustworthy FL to later create a comprehensive taxonomy with the most relevant ones. The proposed taxonomy extends the pillars already identified by previous work (privacy, robustness, fairness, explainability, and accountability) with a new one called federation. This new pillar quantifies the trustworthiness of FL models from the participants and FL model perspectives. Furthermore, those pillars identified by the literature have also been extended with different notions and new metrics dealing with FL models. Then, based on the proposed taxonomy, a trustworthiness evaluation algorithm for FL models, named \textit{FederatedTrust}, has been designed to be extensible, configurable, and flexible. One of the most relevant and powerful FL frameworks, FederatedScope, was selected as the reference FL framework to deploy and test the prototypical implementation of FederatedTrust. Four experiments were performed to validate the FederatedTrust prototype with different FL settings varying the number of clients, training rounds, and differential privacy parameters. For each experiment, the trustworthiness level of each pillar and metric were compared and discussed while solving the problem of classifying handwritten digits (using the FEMNIST dataset) in a collaborative and privacy-preserving way. The experiments demonstrated the complexity of the task of quantifying the trustworthiness level of FL models, being the FederatedTrust algorithm the first attempt to holistically assess an FL model based on a comprehensive trustworthiness taxonomy. To conclude, existing limitations of the current version of the algorithm prototype have also been discussed.}{This work presents a comprehensive taxonomy encompassing the most relevant aspects for trustworthy FL. The proposed taxonomy expands upon the existing pillars previously identified by prior research, namely privacy, robustness, fairness, explainability, and accountability, by introducing a novel on called federation. This new pillar quantifies the trustworthiness of FL models from both participant and FL model perspectives. Moreover, the existing pillars recognized in the literature have been augmented with various notions and novel metrics that address FL models. Based on the proposed taxonomy, a trustworthiness evaluation algorithm for FL models, named FederatedTrust, has been devised with the aim of being extensible, configurable, and flexible. To assess the effectiveness and viability of the FederatedTrust prototype, it was implemented and tested within the FederatedScope FL framework. Five experiments were conducted to validate the FederatedTrust prototype, utilizing distinct FL configurations that involved varying the number of clients, datasets, training rounds, differential privacy parameters, normalization and data balancing approaches, and model configurations. Throughout these experiments, the trustworthiness levels of each pillar and metric were compared and analyzed while addressing the task of classifying handwritten digits using the FEMNIST dataset, and detecting IoT malware using the N-BaIoT dataset. The experiments demonstrated the intricate nature of quantifying the trustworthiness level of FL models, with the FederatedTrust algorithm representing the initial endeavor to comprehensively assess an FL model based on a holistic trustworthiness taxonomy. Additionally, existing limitations of the current version of the algorithm prototype have been discussed.}

As future work, it is planned to extend the prototype by implementing new metrics identified in the taxonomy. It is also intended to deploy the algorithm on FL frameworks training FL models in a decentralized fashion. \change{Furthermore, the score aggregation step could be improved by analyzing the trade-offs between pillars and metrics to handle the weight assignments. Finally, the normalization functions could also be improved by designing better score ranges and mappings.}{Furthermore, multi-objective optimization techniques will be analyzed and evaluated to help in the the task of aggregating metrics while computing the global trust score.}



\section*{ACKNOWLEDGMENT}


This work has been partially supported by \textit{(a)} the Swiss Federal Office for Defense Procurement (armasuisse) with the DEFENDIS and CyberForce (CYD-C-2020003)  projects and \textit{(b)} the University of Zürich UZH.


\bibliographystyle{unsrt}
\bibliography{references}

\end{document}